\documentclass{aa}
\usepackage[varg]{txfonts}
\usepackage{graphicx}
\usepackage{dblfloatfix}
\usepackage{epstopdf}
\usepackage{pdflscape}
\usepackage{booktabs}
\usepackage{txfonts}
\usepackage{gensymb}
\usepackage{color}
\usepackage{url}
\usepackage{multirow}
\usepackage{amsmath}
\usepackage{amstext}
\usepackage{natbib}
\usepackage{longtable}
\usepackage{float}
\usepackage[export]{adjustbox}

\begin{document}

\title{Probing the surface environment of large T-type asteroids}

\author{Yuna G. Kwon\inst{\ref{inst1}\thanks{Alexander von Humboldt Postdoctoral Fellow}}~\and Sunao Hasegawa\inst{\ref{inst2}}~\and Sonia Fornasier\inst{\ref{inst3},\ref{inst4}}~\and Masateru Ishiguro\inst{\ref{inst5},\ref{inst6}}~\and Jessica Agarwal\inst{\ref{inst1},\ref{inst7}}
 }

\institute{Institut f{\" u}r Geophysik und Extraterrestrische Physik, Technische Universit{\" a}t Braunschweig,  Mendelssohnstr. 3, 38106 Braunschweig, Germany (\email{y.kwon@tu-braunschweig.de}\label{inst1})
\and Institute of Space and Astronautical Science, Japan Aerospace Exploration Agency, 3-1-1 Yoshinodai, Chuo-ku, Sagamihara, 252-5210 Kanagawa, Japan\label{inst2}
\and LESIA, Universit{\'e} Paris Cit{\'e}, Observatoire de Paris, Universit{\'e} PSL, Sorbonne Universit{\'e}, CNRS, 92190 Meudon, France\label{inst3}
\and Institut Universitaire de France (IUF), 1 rue Descartes, 75231 Paris cedex 05, France\label{inst4}
\and Department of Physics and Astronomy, Seoul National University, 1 Gwanak-ro, Gwanak-gu, 08826 Seoul, Republic of Korea\label{inst5}
\and SNU Astronomy Research Center, Seoul National University, 1 Gwanak-ro, Gwanak-gu, 08826 Seoul, Republic of Korea\label{inst6}
\and Max Planck Institute for Solar System Research, Justus-von-Liebig-Weg 3, 37077 G{\"o}ttingen, Germany\label{inst7}
}

\date{Received \today / Accepted ---}

\abstract {The thermal and radiative environments asteroids encountered have shaped their surface features. Recent observations have focused on asteroids in the main belt showing implications for ices and organics in their interiors likely significant on prebiotic Earth. They stand out in reflectance spectra as exhibiting darker, redder colours than most colocating asteroids.}
{We aim to probe the surface environment of large ($>$80 km in diameter) T-type asteroids, a taxonomic type relatively ill-constrained as an independent group, and therefrom discuss their place of origin.}
{We performed spectroscopic observations of two T-type asteroids, (96) Aegle and (570) Kythera, over the $L$-band (2.8--4.0 $\mu$m) using the Subaru telescope. With other T-type asteroids' spectra available in the literature and survey datasets, we strove to find commonalities and global trends in this group. We also utilised the asteroids' archival polarimetric data and meteorite spectra from laboratory experiments to constrain their surface texture and composition.}
{Our targets exhibit red $L$-band continuum slopes, (0.30$\pm$0.04) $\mu$m$^{\rm -1}$ for (96) Aegle and (0.31$\pm$0.03) $\mu$m$^{\rm -1}$ for (570) Kythera, similar to (1) Ceres and 67P/Churyumov-Gerasimenko, and have an OH-absorption feature with band centres $<$2.8 $\mu$m. 
(96) Aegle hints at a shallow N--H band near 3.1 $\mu$m and C--H band of organic materials over 3.4--3.6 $\mu$m, whereas no diagnostic bands of water ice and other volatiles exceeding the noise of the data were seen for both asteroids. The large T-type asteroids but (596) Scheila display similar spectral shapes to our targets. $\sim$50 \% of large T-types contain an absorption band near 0.6--0.65 $\mu$m likely associated with hydrated minerals. For T-type asteroids (except Jupiter Trojans) of all sizes, we found a weak correlation that the smaller the diameter and the closer the Sun, the redder the visible (0.5--0.8 $\mu$m) slope.}
{The 2.9-$\mu$m band depths of large T-types suggest that they might have experienced aqueous alteration comparable to Ch-types but more intense than most of the main-belt asteroids. The polarimetric phase curve of the T-types is well described by a particular surface structure. The 0.5--4.0 $\mu$m reflectance spectra of large T-type asteroids appear most similar to CI chondrites with grain sizes of $\sim$25--35 $\mu$m. Taken as a whole, we propose that large T-type asteroids might be dislodged roughly around 10 au in the early solar system.}


\titlerunning{Spectroscopic study of large T-type asteroids}

\authorrunning{Y. G. Kwon et al.}

\maketitle

\section{Introduction \label{sec:intro}}

As leftovers from the planet formation process, asteroids occupy distinct regions relative to the Sun as a result of perpetual gravitational scattering by planets and mutual collisions. Their wide range of surface reflectances and colours enables researchers to establish taxonomical systems by grouping asteroids showing similarities in spectral features (e.g. \citealt{Tholen1984,Bus2002,DeMeo2009}). Asteroids in the same taxon often serve as the basis for understanding the formation and dynamical evolution of a certain region in the solar nebula \citep{DeMeo2014}.

\begin{figure}[!t]
\centering
\includegraphics[width=9cm]{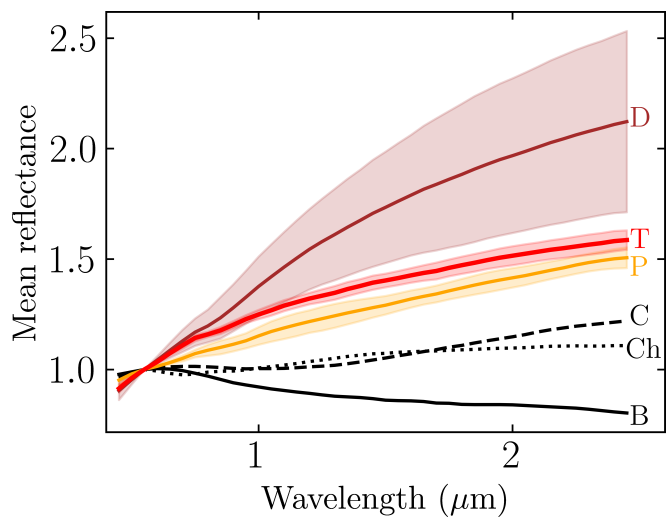}
\caption{Mean reflectance of D-, T-, C-, Ch, and B-type asteroids normalised at 0.55 $\mu$m from the Bus-DeMeo taxonomy \citep{DeMeo2009}. We also included the P-type from the former Tholen taxonomy, which falls in the X-complex in \citet{DeMeo2009}. Among various sub-types in this complex (e.g. Xc, Xe, Xk, and so on), we plotted mean reflectance of X-type asteroids (`X\_Mean' on offer) as a representative of the P-type. The shaded areas of top three curves denote the standard deviations of each taxon. 
} 
\label{Fig01}
\end{figure}

Of dozens of spectral types, D- and T-type asteroids stand out in their redder, featureless spectra over 0.45--2.45 $\mu$m (VNIR) than most colocating main-belt asteroids (Fig. \ref{Fig01}). These very-red dark (albedo $<$ 0.1) asteroids are thought to retain abundant primitive materials inside, such as organics and ice  \citep{Vernazza2015}. Compared to D-type asteroids that have been actively studied (e.g. \citealt{Emery2004,Brown2016}), T-type asteroids are minor{\footnote{23 as of March 2022, labelled as T either by Tholen or Bus-DeMeo taxonomies \citep{Tholen1984,DeMeo2009}. Among them 7 asteroids reside outside of the main-asteroid belt.}}, and no consensus on their origin has yet been established. \cite{Britt1992} suggested a possible relation between T- and M-type asteroids due to their troilite-like spectra and similar semi-major axes, while the sharp OH-band shape of T-type (308) Polyxo in the so-called 3-$\mu$m region (2.8--3.2 $\mu$m), like those of B-type (2) Pallas and C-type (54) Alexandra, implies that T-type asteroids might have experienced aqueous alteration once prevalent across the mid-asteroid belt \citep{Rivkin2002,Takir2012,Takir2015}. Meanwhile, similar VNIR spectra between T-type asteroids and volatile-rich meteorites suggest their link to the outer solar system
\citep{Hiroi2003,Hiroi2005}. Given that T-type asteroids are often omitted from analysis or lumped into other types, at this juncture, conducting in-depth research focusing only on T-type asteroids is worthwhile in adding finer details to the evolution of our solar system.

The physical and compositional properties of asteroids' surfaces can be studied via reflectance spectroscopy. In the VNIR region, overtones and combinations of the functional groups of H$_{\rm 2}$O/OH and hydrocarbons exist, but small amounts of mixtures of opaque elements (e.g. amorphous carbon and magnetite) and weathering effects can easily mute the weaker absorption features of interest \citep{Clark1981,Cruikshank1992,Encrenaz2008}, making it challenging to diagnose the surface status for T-like featureless spectra. In contrast, stronger fundamental bands of interest than their overtones in the VNIR emerge at longer near-infrared wavelengths (LNIR; 2.5--4.0 $\mu$m) \citep{Gaffey1993}. For instance, absorption features in the 3-$\mu$m region are near the local maximum of the absorption coefficients of hydrous materials (i.e., minerals including H$_{\rm 2}$O or OH), such that this region has been actively investigated to enlighten the thermal history of the early solar system \citep{Takir2012,Usui2019}. The N--H, C--H, and C--O bands exist across the LNIR region indicating the presence of phyllosilicates, ices, hydrocarbons, and salts, allowing constraints on the temperature at asteroid-forming epochs and subsequent evolution  \citep{Rivkin2002,Clark2007,DeSanctis2018,Kurokawa2020}.

We present new LNIR spectroscopic observations of large T-type asteroids, (96) Aegle and (570) Kythera, obtained from the Subaru Telescope. Their large size ($\sim$178 km and $\sim$88 km in diameter for the former and the latter, respectively{\footnote{NASA/JPL Small Body Query Database (\url{https://ssd.jpl.nasa.gov/tools/sbdb\_query.html})}}) would likely keep them intact from catastrophic collisions \citep{Bottke2005,Charnoz2007}. Combining our results with published LNIR spectra and survey data of other large T-type asteroids, we examined their surfaces' texture and grain composition and therefrom discussed a plausible scenario for their formation and subsequent dynamical evolution. We here regard a T-type asteroid (T-type, for short) as a body ever labelled as T either by the Tholen \citep{Tholen1984} or Bus-DeMeo \citep{DeMeo2009} taxonomy. Sect. \ref{sec:obsdata} describes the observational methods and data analyses, and in Sect. \ref{sec:res}, we present the results, which are discussed in Sect. \ref{sec:discuss}. All the approaches will be wrapped together into our conclusions in Sect. \ref{sec:sum}.
\\

\begin{table*}[!h]
\centering
\caption{Observational geometry and instrument settings}
\vskip-1ex
\begin{tabular}{c|c|c|c|c|c|ccccc}
\toprule
Telescope/ & Median UT & \multirow{2}{*}{Object} & \multirow{2}{*}{$N$} & {Exptime} & \multirow{2}{*}{$X$} & $m_{\rm V}$ & $r_{\rm H}$ & $\Delta$ & $\alpha$ & Standard\\
Instrument & (2020 Sep 19) & & & (sec) &  & (mag) & (au) & (au) & (\degree) & star\\
\midrule
\midrule
\multirow{5}{*}{Subaru/IRCS} & \multirow{2}{*}{11:59:02} & \multirow{2}{*}{(96) Aegle} & \multirow{2}{*}{4} & \multirow{2}{*}{1 920} & 1.12 & \multirow{2}{*}{12.862} & \multirow{2}{*}{3.460} & \multirow{2}{*}{2.490} & \multirow{2}{*}{5.10} & \multirow{2}{*}{HD 12846} \\
 & & & & & (1.02--1.21) & & & & & \\
\cmidrule{2-11}
 & \multirow{2}{*}{13:50:58} & \multirow{2}{*}{(570) Kythera} & \multirow{2}{*}{5} & \multirow{2}{*}{2 400} & 1.13 & \multirow{2}{*}{13.436} & \multirow{2}{*}{3.016} & \multirow{2}{*}{2.112} & \multirow{2}{*}{9.97} & \multirow{2}{*}{HD 377} \\
 & & & & & (1.04--1.22) & & & & & \\
\bottomrule
\end{tabular}
\tablefoot{Top headers: $N$, number of one dither set of exposures; Exptime, the total integration time in seconds, which is a product of the exposure time per frame (EXP1TIME in the FITS header), the number of nondestructive reads (NDR), and the number of the exposure at each slit position (COADD) by the total number of the asteroid spectra combined in each set (4$N$); $X$, mean airmass with the range in airmass in the bracket; $m_{\rm V}$, apparent $V$-band magnitude provided by the NASA/JPL Horizons system (http://ssd.jpl.nasa.gov/?horizons); $r_{\rm H}$ and $\Delta$, median heliocentric and geocentric distances in au, respectively; $\alpha$, median phase angle (angle of Sun--comet--observer) in degrees.} 
\label{t1}
\vskip-1ex
\end{table*}

\section{Observations and data analysis \label{sec:obsdata}}

In this section, we describe the observational circumstances and data analyses. The journal of observational geometry and instrument settings are summarised in Table \ref{t1}.

\subsection{Observations \label{sec:obs}}

2.8--4.2 $\mu$m spectroscopic observations of (96) Aegle and (570) Kythera were conducted on UT 2020 September 19 using the Infrared Camera and Spectrograph (IRCS; \citealt{Kobayashi2000}) mounted on the Nasmyth focus of the 8.2-m Subaru Telescope (155\degr28$\arcmin$34$\arcsec$W, 19\degr49$\arcmin$32$\arcsec$N, 4 139 m) atop Mauna Kea, Hawaii. Low-resolution ($\Delta\lambda$/$\lambda$ $\sim$ 150; pixel scale of 52 mas/pixel) grism spectroscopy mode with adaptive optics (AO188; \citealt{Hayano2010}) was implemented with a slit width of 0\farcs3 for both asteroids and standard stars (solar analogues of G2V spectral type). The CFHT Sky Probe monitor showed that the night of observation was photometric, and AO188 enabled to reach a diffraction-limited full-width-at-half-maximum down to 0\farcs12. 

To subtract the sky emissions and detector's dark current, object spectra were taken at two positions (A and B) on the detector, separated by 4$\arcsec$ along the slit direction, and nodded in the A$-$B$-$B$-$A dither cadence. We repeated this procedure $\ge$4 times for each asteroid to obtain a high signal-to-noise ratio for combined spectra, achieving total integration times of 1 920 for (96) Aegle and 2 400 s for (570) Kythera. The total number of the spectral images is 4 $\times$ the number of A$-$B$-$B$-$A cadences, that is, 4$N$ in Table \ref{t1}. The telescope was tracked to follow the targets' non-sidereal motions during which observations of standard stars were interspersed at airmass similar to the target's (different by $\lesssim$0.02 on average). The observations of both targets were conducted with the same instrumental settings and exposure time per frame. The standard stars and targets were observed at the same slit position that was set along the North--South direction (SLIT\_PA = 90$^{\rm \circ}$). The supporting astronomer confirmed that atmospheric dispersion was negligible. Spectral flat frames were obtained at the end of observations.

\subsection{Data analysis \label{sec:data}}

The procedure basically followed the `Grism Spectroscopic Observations with IRCS (ver. 2.1.3E)'\footnote{\url{https://www.naoj.org/Observing/DataReduction/Cookbooks/IRCSsp_cookbook_2010jan05.pdf}} that offers IRAF and shell software\footnote{\url{https://www.naoj.org/Observing/DataReduction/index.html}} to reduce of the IRCS grism data.
The following paragraphs spell out two topics: (i) the pre- and post-processing of the grism spectroscopic data; and (ii) intercomparison of the standard stars' spectra used to estimate calibration uncertainties.

Firstly, all spectral images of 1 024 $\times$ 1 024 pixels were reduced to subtract the nod positions from each other (A$-$B) to eliminate OH sky emissions. The 4$N$ images of a target were then divided into two groups (i.e., having 2$N$ files), one group taken in the first half and the other taken in the second half of the observation. The images in each group were median-combined into a single spectrum of the target, yielding two target images. The same procedures were applied to the standard stars. Next, we cropped the combined spectral images to leave 340 $\times$ 860 pixels in the x and y directions, respectively, with the object in the centre, and divided them by a normalised flat frame to remove pixel-to-pixel sensitivity variations on the chip. We used differential images of the lamp-on (i.e., the light from the lamp plus thermal emission of the system) and lamp-off (thermal emission only) frames to construct flat-field images. For bad pixel correction, we first co-added all flat images into a single file, divided the file by the median-combined master flat to build a bad-pixel map, and subtracted the map from the target and standard star images. It was confirmed that no bad pixels overlapped with the object positions. The resultant object images were further checked to determine whether there were any ill-behaved pixels showing $\ge$5$\sigma$ higher or lower counts than the mean of the surrounding pixels due to a hit of cosmic rays or pixel malfunctions. If any, the 5 $\times$ 5 neighbouring pixels were used to interpolate the mean value to the affected pixel. 

After preprocessing, we extracted 1-D source spectra (counts as a function of pixels) along the dispersion axis from the 2-D spectral images. The background of each spectrum was fitted by a linear function (i.e., second-order Chebyshev function) with a 3$\sigma$ rejection cut and subtracted from the observed signals. Third-order and second-order Legendre polynomial fitting functions were then applied to extract the spectra  for the standard stars and targets, respectively, yielding the root-mean-squares (RMS) of $\sim$0.03 for the stars and $\sim$0.15 pixels for the targets. The extracted spectra were then wavelength-calibrated by matching the near-infrared sky transmission spectrum at Mauna Kea\footnote{\url{https://www.subarutelescope.org/Observing/Instruments/IRCS/IRCSum_1.0.1.pdf}} to the absorption features of the  standard stars and targets. 
We fit the absorption features versus the x coordinate using a third-order spline3 (cubic spline) function. The RMS of the fits were always an order of $\sim$0.03 \AA\ (pixel)$^{\rm -1}$ and $\sim$0.20 \AA\ (pixel)$^{\rm -1}$ for standard stars and targets, respectively (which are acceptable given that the dispersion is 15.9 \AA\ (pixel)$^{\rm -1}$ in the $L$ band\footnote{\url{https://www.naoj.org/Observing/Instruments/IRCS/grism/grisms.html}}).
The calibrated spectra were binned three times to reduce the sampling cadence ($\sim$1200) to the spectral resolution ($\sim$150), where the data at $<$2.85 $\mu$m and $>$4.10 $\mu$m were discarded due to the nonlinearity of the detector in intensity and intrinsically small atmospheric transmission thereof (based on the manual's recommendation). The binned spectra were then normalised, after which the two groups of the target/standard star spectra were averaged into a single spectrum.
We divided the resulting spectrum of the target by the spectrum of a standard star whose airmass differs on average by within 0.02. Since the standard stars used are G2V-type solar analogues, we were able to obtain normalised reflectance for each target. 

\begin{figure}[!t]
\centering
\includegraphics[width=8.5cm]{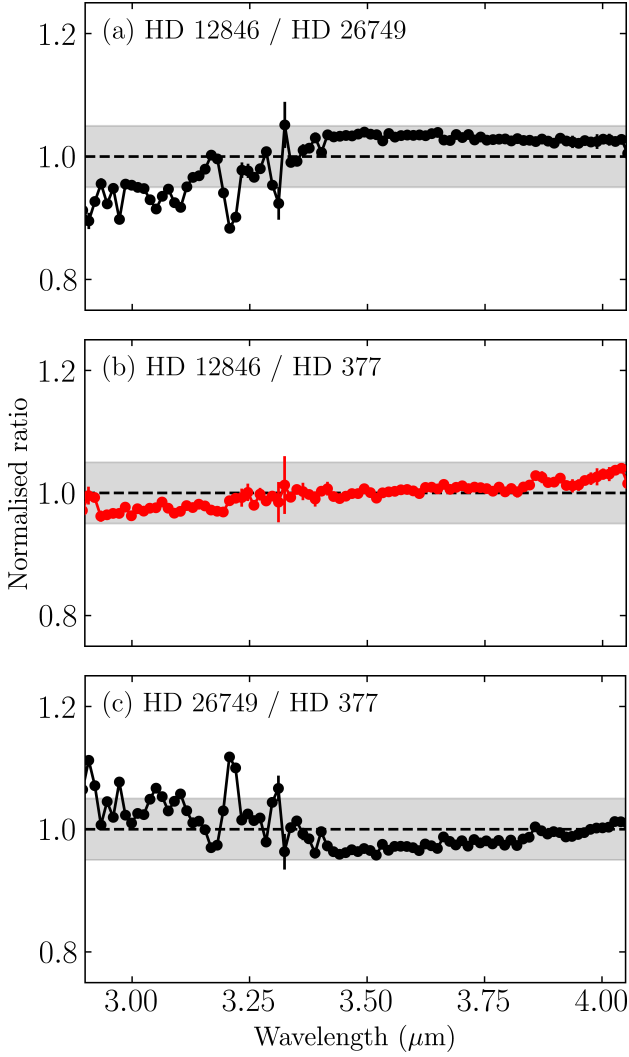}
\caption{Ratios of the normalised counts of the standard stars. The horizontal grey-shaded areas represent $\pm$5 \% uncertainty on unity. We selected HD 12846 and HD 377 for calibration and their normalised ratio is highlighted as red colour in panel b.}
\label{Fig02}
\end{figure}

A systematic difference in the standard stars was measured to estimate the calibration uncertainty in the spectral slope of the targets. We observed a total of three standard stars during the night (G2V-type HD 26749, in addition to the two stars in Table\ref{t1}) and evaluated the consistency of their normalised counts as a function of wavelength by dividing one spectrum by another. Comparing the stellar spectra in Figure \ref{Fig02} justified our selection of standard stars (HD 12846 and HD 377) by showing that their spectral slopes are internally consistent with an accuracy of $\pm$5 \% (grey-coloured areas) on unity over $\sim$2.9--4.0 $\mu$m. This level of uncertainty is comparable to the linearity uncertainty ($\sim$3--5 \%) of the 0\farcs3 slit width in $<$10 000 counts, as described in the official instrument manual. Bumpy structures near 3.3 $\mu$m are due to the atmospheric CH$_{\rm 4}$ absorption \citep{Lord1992}.
Deviations larger than $\pm$5 \% uncertainty level were persistently observed at wavelengths of <2.9 and >4.0 $\mu$m, leading us to ignore these regions in the following analysis.

In summary, we delimited the data used for analysis from 2.9 to 4.0 $\mu$m from the original wavelength coverage sampled by the instrument from 2.8 to 4.2 $\mu$m. A slope of the target continuum has an additional $\pm$5 \% error on its nominal value on account of the calibration uncertainty of the standard stars. In the following sections, the term `$L$ band' refers to this effective wavelength region of 2.9--4.0 $\mu$m and will be interchangeably used with `LNIR (2.5--4.0 $\mu$m).'
\\

\section{Results \label{sec:res}}

We examined the spectral properties of (96) Aegle and (570) Kythera. Other T-types' LNIR spectra from the literature, the second phase of the Small Main-belt Asteroid Spectroscopic
Survey (SMASSII; \citealt{Bus2002}) datasets and its extension to 2.5 $\mu$m \citep{DeMeo2009} were also employed to explore global trends in this taxonomic group. 

\subsection{LNIR spectra of (96) Aegle and (570) Kythera \label{sec:res1}}

\begin{figure}[!t]
\centering
\includegraphics[width=8cm]{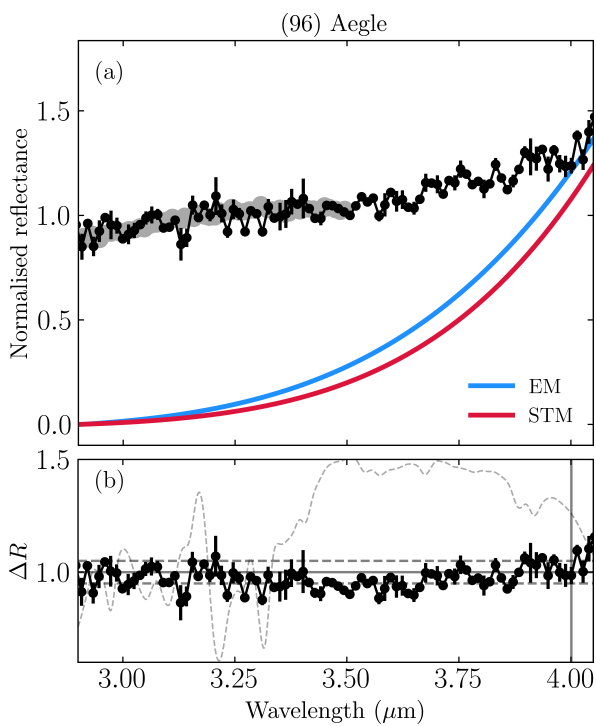}
\caption{Reflectance {\it R} versus wavelength of (96) Aegle, normalised at 3.5 $\mu$m. (a) Background grey circles are quoted from \citet{Fraeman2014} where the same asteroid was observed at $r_{\rm H}$ = 3.1 au. Blue and red curves denote the expected thermal flux derived from the equilibrium model (EM) and the standard thermal model (STM), respectively. The modelled fluxes were used to compare the spectral shapes with the observed spectrum to diagnose the influence of thermal contamination with the assumption that their contribution to the observed flux is negligible at 2.9 $\mu$m. (b) A continuum-removed spectrum ($\Delta R$) is shown with the upper and lower dashed lines of the $\pm$5 \% calibration uncertainty. The dashed curve in the background shows the atmospheric transmission curve modelled by ATRAN \citep{Lord1992} in the same median airmass of (96) Aegle (Table \ref{t1}). We multiplied the curve by 1.6 to compare the atmospheric pattern to the observed spectrum. The solid vertical line marks the shortest wavelength where the contribution of the thermal flux to the observed flux becomes non-negligible.}
\label{Fig03}
\end{figure}

\begin{figure}[!b]
\centering
\includegraphics[width=8cm]{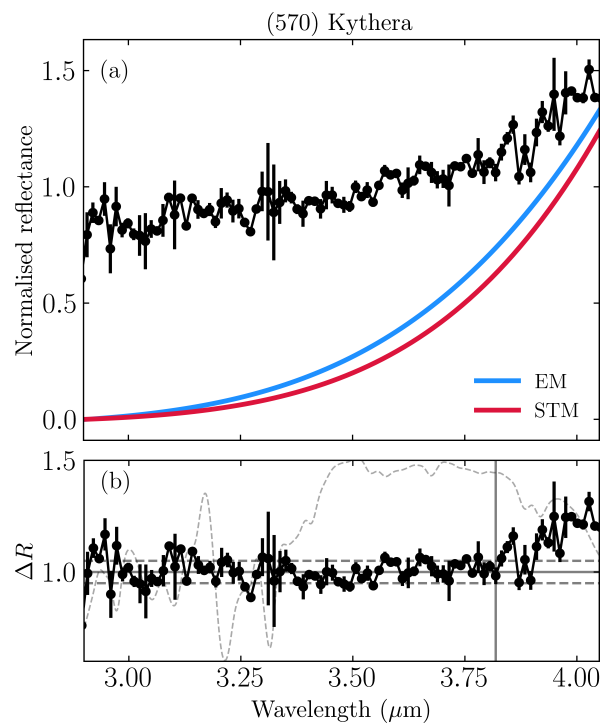}
\caption{Same as Figure \ref{Fig03}, but for (570) Kythera.}
\label{Fig04}
\end{figure}

An intrinsic spectral shape of an asteroid can be expressed as `normalised reflectance ({\it R})' by dividing the asteroid's spectrum by that of a standard star. Given that the data of the normalised target and standard star are represented by $X_{\rm n} \pm \sigma_{\rm X_{\rm n}}$ and $Y_{\rm n} \pm \sigma_{\rm Y_{\rm n}}$, respectively, the propagated uncertainty associated with $R$ of $X_{\rm n}/Y_{\rm n}$ is $\sqrt{\big(\frac{\sigma_{\rm X_{\rm n}}}{Y_{\rm n}}\big)^{\rm 2} + \big(\frac{\sigma_{\rm Y_{\rm n}} X_{\rm n}}{Y_{\rm n}^{\rm 2}}\big)^{\rm 2}}$. The $X_{\rm n}$ and $Y_{\rm n}$ are normalised values of $X$ and $Y$, respectively, which are the average of the two groups' median-combined spectra (Sect. \ref{sec:data}). Figures \ref{Fig03} and \ref{Fig04} show the resulting {\it R} of (96) Aegle and (570) Kythera normalised at 3.5 $\mu$m.  

Overall reflectances of both asteroids increase linearly (i.e., red spectral slopes) except for the long end of the $L$-band region. In general, the contribution of thermal emission to the observed flux for main-belt asteroids becomes significant beyond 3.5 $\mu$m \citep{Lebofsky1981}. We adopted simple thermal models -- the equilibrium model (EM, i.e., a modified blackbody function; \citealt{Delbo2004}) and standard thermal model (STM; \citealt{Lebofsky1986}) -- to check the distribution of thermal components in our observed data. The thermal portion in the EM was calculated using a first-order blackbody function in instantaneous equilibrium with solar insolation:
\begin{equation}
f_{\rm EM}(\lambda) = \frac{\epsilon \pi B(\lambda,T_{\rm SS})R_{\rm e}^{\rm 2}}{\Delta^{\rm 2}}
~,
\label{eq:eq1}
\end{equation}
\noindent where $\epsilon$ is the wavelength-independent emissivity (0.9 used here; \citealt{Lebofsky1980}), $B(\lambda,T_{\rm SS})$ is the Planck radiation function at the sub-solar temperature $T_{\rm SS}$, $R_{\rm e}$ is the effective radius of an asteroid, and $\Delta$ is the geocentric distance at the observing epoch. The STM requires additional steps to consider a distribution in the surface temperature over the sunlit hemisphere of the asteroid $T(\Phi)$ and the effect of surface roughness $\eta$ \citep{Lebofsky1980,Lebofsky1986}:
\begin{equation}
f_{\rm STM}(\lambda) = \frac{\epsilon D_{\rm e}^{\rm 2}}{2\Delta^{\rm 2}} \int_{0}^{\frac{\pi}{2}} B(\lambda,T(\Phi))\sin\Phi\cos\Phi~d\Phi
~,
\label{eq:eq2}
\end{equation}
\noindent where
\begin{equation}
T(\Phi) = \left\{ \begin{array}{ll}
 T_{\rm max}\cos^{\rm 1/4}\Phi & \textrm{if $\Phi \le \pi/2$}\\
 0 & \textrm{otherwise}
\end{array} \right.~,
\label{eq:eq3}
\end{equation}
\noindent and the maximum temperature $T_{\rm max}$ at the sub-solar point was obtained from
\begin{equation}
T_{\rm max} = \Bigg[\frac{(1-A)S}{\epsilon \sigma \eta r_{\rm H}^{\rm 2}}\Bigg]^{\rm \frac{1}{4}}
~.
\label{eq:eq4}
\end{equation}
\noindent Here, $D_{\rm e}$ is the effective diameter of the asteroid, $\Phi$ is the solar colatitude, $S$ is the solar constant (1360.8 W m$^{\rm -2}$), $\sigma$ is the Stefan-Boltzmann constant (5.67 $\times$ 10$^{\rm -8}$ W m$^{\rm -2}$ K$^{\rm -4}$), $\eta$ is the beaming parameter (0.756 for large, low-albedo asteroids; \citealt{Lebofsky1986}), and $r_{\rm H}$ is the heliocentric distance. $A$ is the bolometric Bond albedo and  approximated to that in the $V$ band as $A$ $\approx$ $A_{\rm V}$  = $q$ $\times$ $p_{\rm V}$, where $p_{\rm V}$ is the geometric albedo (0.048 for (96) Aegle\footnote{from NASA/JPL Small-Body Database Query} and 0.044 for (570) Kythera; \citealt{Masiero2011}). $q$ is the phase integral that equates 0.290 + 0.684 $\times$ $G$ in the standard $H$--$G$ magnitude system \citep{Bowell1989}, where $G$ is the slope parameter (0.15 used here). 
In Figures \ref{Fig03}a and \ref{Fig04}a, we overplotted the modelled thermal fluxes assuming their negligible contribution at 2.9 $\mu$m. The modelled curves were used to compare the spectral shapes with the observed spectrum to diagnose the influence of thermal contamination.
The steep rise of thermal components seems along with the observed reflectances at $\gtrsim$4.0 $\mu$m for (96) Aegle and $\gtrsim$ 3.8 $\mu$m for (570) Kythera. The shorter wavelength data show a linear increase and well match the published result of \citet{Fraeman2014} (background grey symbols in Fig. \ref{Fig03}a), justifying negligible slit loss of our data that the implementation of AO188 could achieve. 
Given that thermal inertias of (96) Aegle in 178 km and (570) Kythera in 87 km do not differ much within the measurement errors \citep{Hung2022}, the observed slight difference in the thermal emission of the two asteroids is probably caused by their different heliocentric distances at the time of observation.


We made a least-squares linear fitting over the 2.90--3.75 $\mu$m and measured the spectral slopes of the targets: (0.30$\pm$0.04) $\mu$m$^{\rm -1}$ for (96) Aegle and (0.31$\pm$0.03) $\mu$m$^{\rm -1}$ for (570) Kythera. The continuum was then removed from the observed reflectance of each asteroid (panel b in Figs. \ref{Fig03} and \ref{Fig04}) and the result was compared with the atmospheric transmission curve modelled by ATRAN \citep{Lord1992} in the same airmass of the asteroids to search for plausible absorption features. Most of the data points across the $L$-band fluctuate near the guidelines of the $\pm$5\% calibration uncertainty (Sect. \ref{sec:data}), whereas points of (96) Aegle and (570) Kythera contaminated by the thermal excess deviate larger than the average.
There is a broad, shallow local minimum over 3.4--3.6 $\mu$m on the spectrum of (96) Aegle, which would be compatible with the C--H absorption bands of organic materials \citep{Khare1990}. Its band depth is comparable to but slightly exceeds the $\pm$5 \% uncertainty level. The asteroid also seems to contain an absorption around 3.14 $\mu$m, whose band centre could be associated with the N--H stretch of ammonium ions at $\sim$3.1--3.2 $\mu$m \citep{Lodders2003}. In contrast, the bumpy structures clustered shortward of $\sim$3.2 $\mu$m for (570) Kythera follow similar trends with atmospheric transmittance, primarily attenuated by H$_{\rm 2}$O sky absorptions. A local minimum of $\Delta R$ around 3.25 $\mu$m may be a meaningful feature. However, considering the feature's position in between local maxima of atmospheric absorptions and its neighbouring data points having significant errors due to the CH$_{\rm 4}$ sky absorption at $\sim$3.3 $\mu$m, we suspect that the data points in this region might be affected by the residuals of atmospheric calibration. We thus defer the discussion about this figure to future observations with better data quality. Overall, the reflectance spectra of the two T-types exhibit red and linear continuum slopes in the $L$ band, hinting at the N--H band near 3.1 $\mu$m and C--H band near $\sim$3.5 $\mu$m for (96) Aegle.

\subsection{Large T-type asteroids over the VNIR--LNIR region \label{sec:res2}}

Featureless $L$-band spectra alone are less informative on the surface conditions of the asteroids. We thus extended wavelength coverage to the VNIR region using the SMASSII datasets \citep{DeMeo2009}. Because strong telluric lines and thermal background noise often degrade data quality for LNIR ground-based observations \citep{Cruikshank2001,Emery2003}, only a handful of large T-types have been observed in the $L$-band region: (308) Polyxo \citep{Takir2012}, (596) Scheila \citep{Yang2011}, (773) Irmintraud \citep{Kanno2003}, (96) Aegle (this study and \citealt{Fraeman2014}) and (570) Kythera from this study. We digitised their LNIR spectra from the published figures. The 2.2--3.4 $\mu$m spectrum of (96) Aegle in \citet{Fraeman2014} nicely fits our $L$-band data, the composite spectrum of which was then scaled to the SMASSII spectra. In the case of (570) Kythera, we found no available data concurrently covering the $L$ band and shorter wavelengths to refer to giving an offset.
Instead, we employed its near-infrared albedo ($p_{\rm IR}$) information measured by the NASA Wide-field Infrared Survey Explorer (WISE; \citealt{Wright2010}). A $p_{\rm IR}$ over the W1 (3.4 $\mu$m) and W2 (4.6 $\mu$m) bands contains a sufficient fraction of reflected sunlight, such that the ratio of $p_{\rm IR}$/$p_{\rm V}$ can be used as a measure of their reflectances \citep{Masiero2011,Harris2014}. We thus scaled the observed reflectance of (570) Kythera at 3.4 $\mu$m to 0.0653 ($p_{\rm IR}$) so as to make $p_{\rm IR}$/$p_{\rm V}$ = 0.0653/0.0440 $\approx$ 1.48 \citep{Masiero2011}.  Previous observations for the other three asteroids covered wavelengths in the $K$ (1.8--2.5 $\mu$m) and $L$ bands simultaneously. 

\begin{figure}[!t]
\centering
\includegraphics[width=8cm]{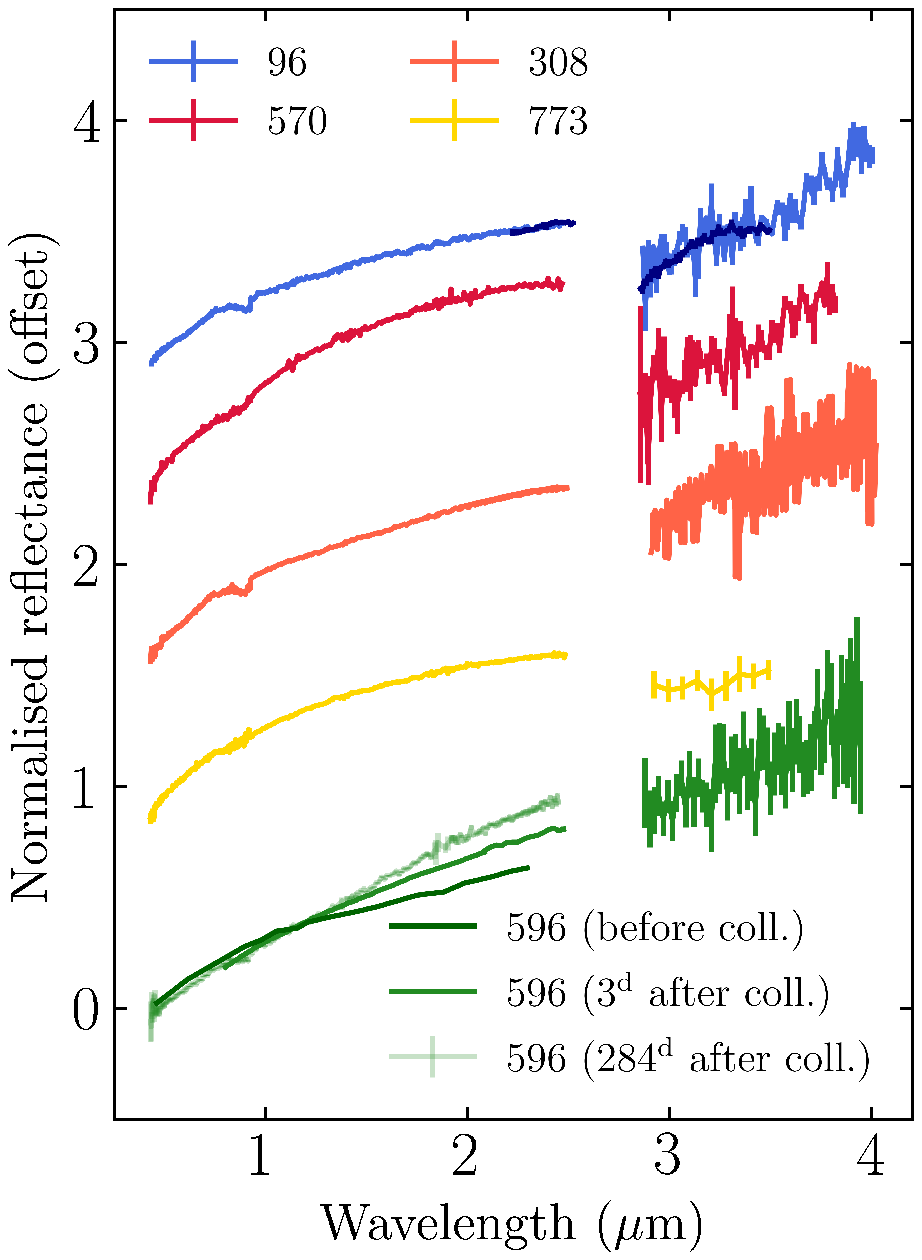}
\caption{Normalised reflectance of large T-type asteroids over 0.4--4.0 $\mu$m. Each was normalised at 0.55 $\mu$m and offset for clarity. The 2.8--4.0 $\mu$m spectra of (96) Aegle and (570) Kythera were from our observations; the 2.2--3.5 $\mu$m spectrum of (96) Aegle (navy line superimposed on the blue curve) was from \citet{Fraeman2014}; The 0.4--4.0 $\mu$m spectrum of (308) Polyxo was from \citet{Takir2012}; The 0.8--4.0 $\mu$m spectrum of (596) Scheila taken three days after its 2010 impact event was from \citet{Yang2011}; The 2.1--3.5 $\mu$m spectrum of (773) Irmintraud was from \citet{Kanno2003}. All VNIR spectra but (308) Polyxo were quoted from the SMASSII datasets \citep{DeMeo2009}. The gap over 2.5--2.8 $\mu$m illustrates the region opaqued by water vapour in Earth's atmosphere. Weak features around 0.9 $\mu$m in all spectra is near the boundary of the visible and near-infrared portions of the spectrum taken at different times from different instruments at edge wavelengths where the detectors are less sensitive \citep{DeMeo2009}.}
\label{Fig05}
\end{figure}

Figure \ref{Fig05} reveals that the large T-type asteroids considered have in common, particularly the slope difference and a reflectance drop-off between the $K$ and $L$ bands. Their VNIR shapes are concaving down, whereas those in the $L$ band are nearly linear. Such discontinuities on either side of the region opaqued by water vapour in Earth's atmosphere indicate the presence of an absorption band. Indeed, their linear increase in reflectance over 2.9--3.2 $\mu$m is a characteristic feature of the so-called  `sharp' or `Pallas' 3-$\mu$m spectral group of asteroids with an OH-band of phyllosilicates as a product of aqueous alteration \citep{Takir2012,Rivkin2015b}. Although we adjusted the $L$-band reflectance level of (570) Kythera using its $p_{\rm IR}$ in the absence of simultaneous coverage of the VNIR and LNIR regions, its discernible difference in the slope curvature between the $K$ and $L$ bands would support the likelihood of a similar absorption band. 

(96) Aegle has a shallow but broad absorption band between 3.4 $\mu$m and 3.6 $\mu$m that has also been suggested for (308) Polyxo \citep{Takir2015}. C--H stretching bands of carbonaceous materials with low C/H ratio occupy similar spectral regions to this feature \citep{Khare1993}. Carbonates also have overtone and combinations of $-$CO$_{\rm 3}$$^{\rm -2}$ in this region \citep{Clark2007} but the absence of relevant absorptions at other wavelengths (e.g. at $\sim$3.9 $\mu$m) for the sharp 3-$\mu$m group asteroids makes this component less likely \citep{Takir2015}. If we take the 3.1-$\mu$m N--H feature of (96) Aegle as a valid signal, previous studies on the large T-type asteroids have yet to report the analogous structure. (596) Scheila exhibits a slightly different shape from other T-type asteroids. Its spectral curvature has changed since the impact in 2010 \citep{Larson2010} in the VNIR region, from the T-like concavity before the event to the D-like steep linearity. Modelling of the spectrum obtained right after the impact restricts the abundance of hydrous minerals or water ice to no more than a few percent \citep{Yang2011}, albeit the exposure of the less-evolved inner surface layers \citep{Ishiguro2011}. We consider this asteroid a special case and will discuss it more in detail in Section \ref{sec:discuss}. 

In summary, large T-types considered here have common characteristics in that 1) their VNIR spectra are red and concaving down toward longer wavelengths; 2) their LNIR spectra are red and have linear slopes; 3) their spectral shape and brightness change discontinuously between the $K$ and $L$ bands; and 4) all large T-types used to compare but (596) Scheila are consistent with having an OH-absorption band, which makes them belong to the sharp 3-$\mu$m group defined by \citet{Takir2012}.

\subsection{Search for correlations \label{sec:res3}}

We enlarged the scope of consideration to T-type asteroids of all sizes by exploiting the SMASSII data of 21 T-types ($\sim$0.9--177.8 km in diameter for T-types constrained their size) but Jupiter Trojans. We strove to find global trends in the T-types and correlations between their spectral characteristics (reflectance, spectral slope, and spectral curvature) with sizes and orbital elements implying any universal physical processes acting on their surfaces. 

\begin{figure}[!t]
\centering
\includegraphics[width=9cm]{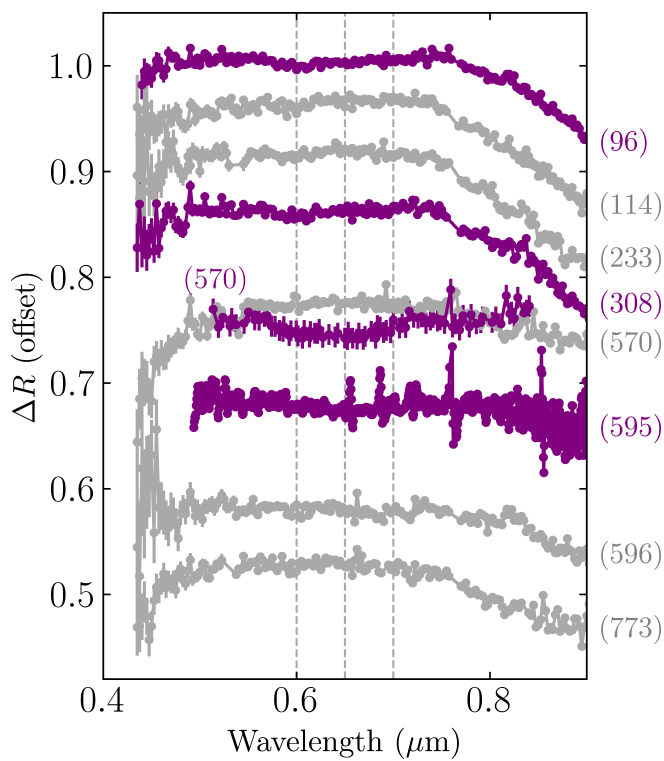}
\caption{Continuum-removed spectra of all currently known large T-types (Table \ref{ta1}). The continuum was fitted by a linear function using data points at $\sim$0.44--0.55 and $\sim$0.72--0.76 $\mu$m (exact start and end points of the wavelengths were adjusted depending on the asteroids' spectral curvature). Each was normalised at 0.55 $\mu$m and offset for clarity. Positive and negative detections of the 0.7 feature were coloured in purple and grey, respectively.}
\label{Fig06}
\end{figure}

One of the widely-used proxies for the hydration status of the surfaces is the so-called 0.7-$\mu$m band (0.6--0.75 $\mu$m), likely attributed to iron oxides in phyllosilicates, a product of aqueous alteration \citep{Vilas1989}. The band centre depends on the mineralogy, specifically cations gluing silica layers of phyllosilicates. For instance, the 0.7-$\mu$m band shifts to shorter wavelengths as the composition becomes more Mg-rich \citep{Cloutis2011b}. Serpentine and smectite group (saponite) phyllosilicates yield a band centre near 0.7-$\mu$m \citep{Vilas1994} and $\sim$0.6--0.65 $\mu$m \citep{Cloutis2011a}, respectively. Intrigued by (570) Kythera showing this absorption band near 0.65 $\mu$m \citep{Vilas1989}, we took a closer look at this spectral region of other large T-types (a total of 8 with $>$80 km in diameter),adding (595) Polyxena that was not initially observed by \citet{Bus2002} but later by \citet{Lazzaro2004} since it shows a visible slope compatible with T-types in the Bus-DeMeo taxonomy. Figure \ref{Fig06} shows their continuum-removed spectra ($\Delta$$R$). A continuum was defined as a linear function fitting the data points over $\sim$0.44--0.55 $\mu$m and $\sim$0.72--0.76 $\mu$m, the exact start and end wavelength points of which were slightly adjusted for each asteroid due to their different spectral curvature. Spectra showing a potential 0.7-$\mu$m feature were highlighted in purple, while ones that did not contain it were shown in grey.

Among eight large T-types, a weak absorption feature appears for (96) Aegle, (308) Polyxo, (570) Kythera, and (595) Polyxena between $\sim$0.60 and 0.65 $\mu$m (highlighted in purple colour). As to (570) Kythera, the SMASSII spectrum (grey line) hints at no feature, whereas that from the NASA/PDS Vilas Asteroid Spectra (\citealt{Vilas2020}, initially observed by \citealt{Vilas1989}) has an absorption band (black line) as the authors reported. Once detected, the band centres of the large T-types appear at shorter wavelengths than those of most hydrous main-belt asteroids (e.g. Ch-type) near $\sim$0.7 $\mu$m \citep{Rivkin2002}. Shallow band depths (at most $\sim$3 \% of the ambient continuum levels) agree with previous observations that the 0.7-$\mu$m band intensity is typically $\sim$1--6 \% of the continuum \citep{Vilas1989,Vilas1994,Fornasier1999,Fornasier2014}. We could not find an equivalent spectral feature for the other four large asteroids ((114) Kassandra, (233) Asterope, (596) Scheila, and (773) Irmintraud), either due to the absence of the materials on the observed surfaces and/or their low content not enough to be detected. For smaller ($<$50 km in diameter) T-types, the fluctuation of the spectra is too large to investigate such subtleties. 

\begin{figure}[!t]
\centering
\includegraphics[width=9cm]{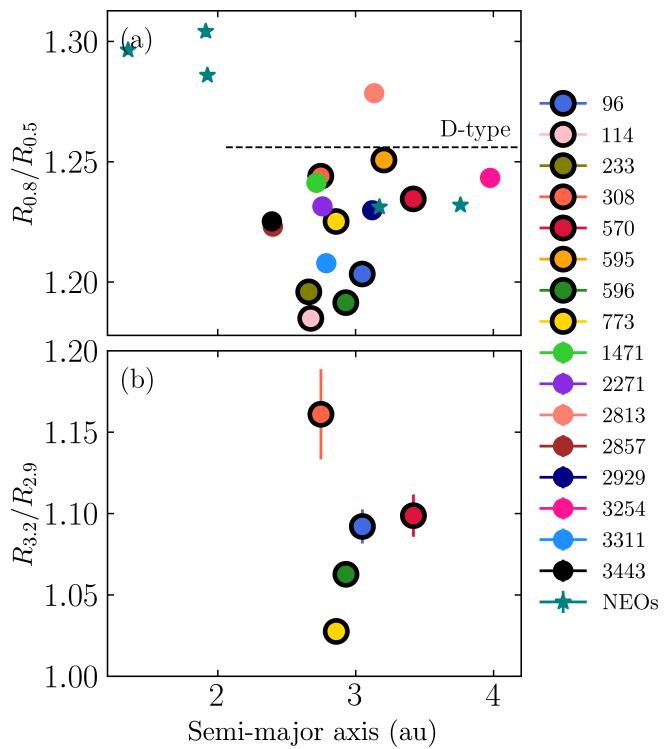}
\caption{Distribution of the reflectance ratios of T-type asteroids as a function of semi-major axes. The ratios of the reflectances measured at 0.8 $\mu$m and 0.5 $\mu$m ($R_{\rm 0.8}/R_{\rm 0.5}$) and at 3.2 $\mu$m and 2.9 $\mu$m ($R_{\rm 3.2}/R_{\rm 2.9}$) are shown in panels (a) and (b), respectively. Circles with thick black edges denote asteroids whose diameter is greater than 80 km, and circles without edges mark smaller ones ($<$50 km). Stars represent five T-types in near-Earth orbits: (85490) 1997 SE5, (138911) 2001 AE2, (162998) 2001 SK162, 2001 UU92, and 2001 YE1. The horizontal line in panel (a) denotes the $R_{\rm 0.8}/R_{\rm 0.5}$ of the D-type's mean reflectance (\citealt{DeMeo2009}; Fig. \ref{Fig01}).}
\label{Fig07}
\end{figure}

Finally, we looked into the distribution of the reflectance ratios (corresponding to the spectral slope) in the VNIR and LNIR regions as a function of semi-major axes. The ratio of the reflectances measured at 0.8 and 0.5 $\mu$m ($R_{\rm 0.8}/R_{\rm 0.5}$) represents the VNIR slope, while that measured at 3.2 and 2.9 $\mu$m ($R_{\rm 3.2}/R_{\rm 2.9}$) represents the LNIR slope. Figure \ref{Fig07} illustrates the results, and all information of the T-type asteroids used in this figure is summarised in Table \ref{ta1}. Symbols were marked depending on the size of the asteroids: circles with thick black edges denote asteroids whose diameter is greater than 80 km; circles without edges indicate smaller asteroids ($<$50 km); and stars denote five T-types whose dynamical properties classify them as near-Earth objects (NEOs). 

Firstly for large T-types compared in Figure \ref{Fig05}, the VNIR and LNIR slopes showcase no apparent relationships with their sizes, albedos, and orbital elements. (We also tested relations with their eccentricity, inclination, and perihelion distance, though not plotted here.) However, when considering T-type asteroids in all size ranges, NEOs tend to have the reddest VNIR slope among the T-types ($R_{\rm 0.8}/R_{\rm 0.5}$ $\sim$ 1.326 on average), followed by intermediate-sized ($\sim$1.223 on average) and larger ($>$80 km) asteroids for the shallowest ($\sim$1.215 on average). T-type NEOs inside 2 au exhibit redder slopes than the NEOs with main-belt semi-major axes. This apparent relationship may be caused by a difference in phase angles during observation. The spectral slopes of D-like red asteroids and comet nuclei have been found to change with phase angle, where  they generally become redder with increasing phase angle \citep{Fornasier2016,Lantz2018}. To see whether this phase reddening shapes the colour distribution, we checked the phase angle of each data point and confirmed that phase angles of the asteroids distribute randomly, spanning from 0\fdg8 to 23\fdg4 except for one NEO (2001 UU92) observed at the phase angle of 59\fdg8 \citep{Binzel2004}. The phase angles of the other NEOs are in the range of the main-belt T-types' phase angles. Given 1) the negligible correlation between the phase angles and $R_{\rm 0.8}/R_{\rm 0.5}$ values (Spearman's correlation $r$ = $-$0.029 for the asteroids analysed here), and 2) the reddening in this phase angle range for dark main-belt asteroids, usually less significant than the observational uncertainties \citep{Marsset2020,Beck2021a}, the result might not be associated with the phase angle conditions of the asteroids. 

Apart from the moderate connection between the spectral slope and diameter of the asteroids, we found no further correlations between the T-type asteroids' properties and their dynamical characteristics, let alone between the slope and albedo of the asteroids ($r$ = 0.002), within the limited number of T-type asteroid data currently available. 
Such an absence of global correlations seems in line with a recent survey of the larger number of low-albedo asteroids by \citet{Usui2019}. 
\\

\section{Discussion \label{sec:discuss}}

In this section, we discuss the surface environment of large T-type asteroids based on the results in Sect. \ref{sec:res} by comparing our findings with previous observations of low-albedo asteroids in other taxonomic types and the nucleus of 67P/Churyumov-Gerasimenko and laboratory experiments on carbonaceous chondrites. We constrain large T-types' hydration status (Sect. \ref{sec:dis1}), surface texture (Sect. \ref{sec:dis2}), and plausible environmental conditions at the accretion epoch (Sect. \ref{sec:dis3}). Based on all evidence currently at our disposal, we make a tentative conclusion in Sect. \ref{sec:dis4} on their place of origin. 

\subsection{Hydration status of large T-type asteroids \label{sec:dis1}}

\begin{figure}[!b]
\centering
\includegraphics[width=9cm]{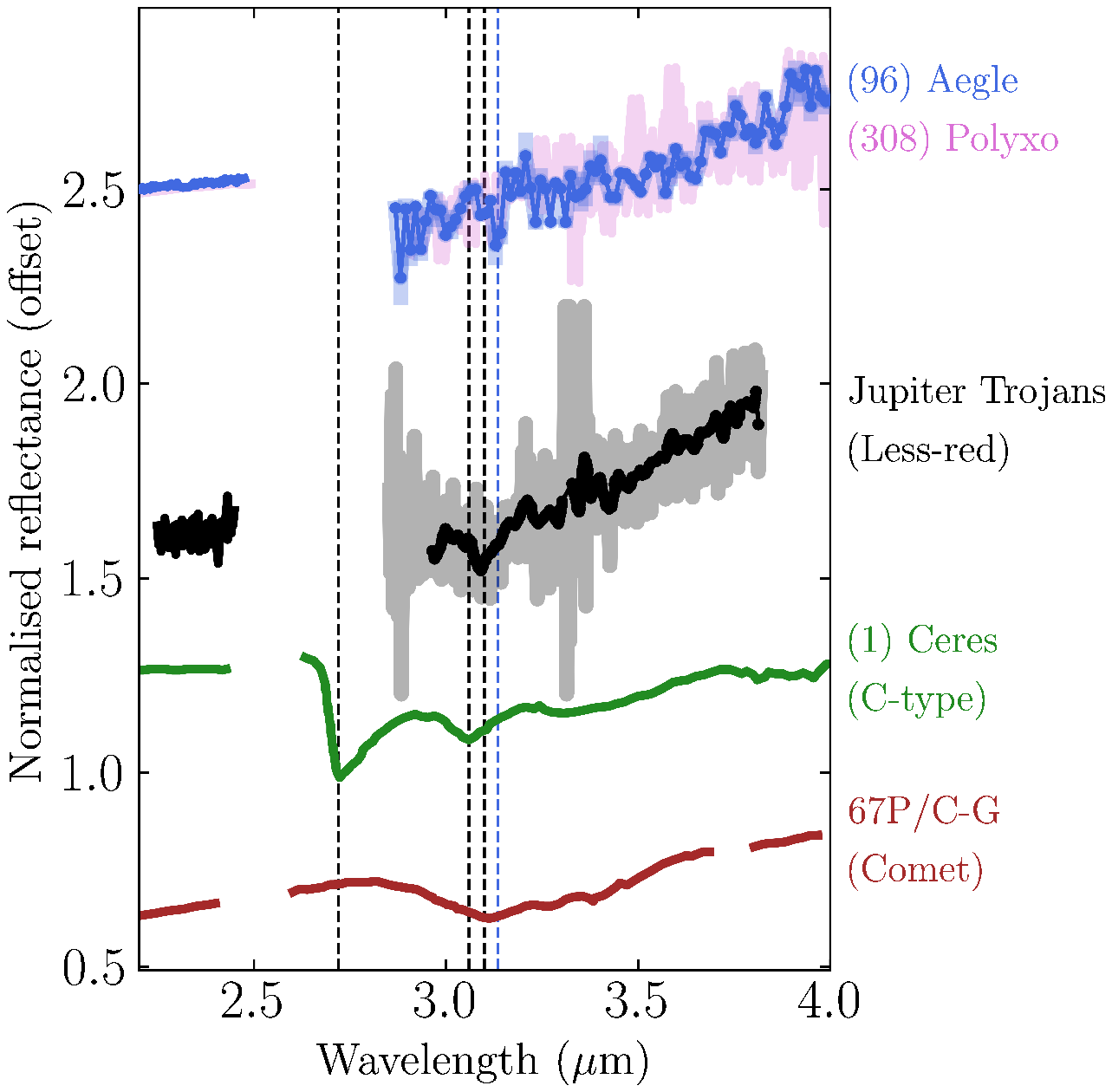}
\caption{Comparison of the LNIR spectra of large T-type asteroids with low-albedo asteroids in different taxonomic types and comet 67P/C-G. (96) Aegle and (308) Polyxo \citep{Takir2012} represent large T-types and are compared with the spectra of Jupiter Trojans \citep{Brown2016}, (1) Ceres \citep{DeSanctis2018}, and comet 67P/C-G \citep{Poch2020}. Each spectrum was arbitrarily offset for clarity. Four vertical lines indicate the absorption bands of 2.72 $\mu$m, 3.06 $\mu$m, 3.1 $\mu$m, and 3.14 $\mu$m. The first (OH) and second (N--H) bands were spotted on (1) Ceres \citep{Carrozzo2018,DeSanctis2018}, the third (N--H) band was found on Jupiter Trojans \citep{Brown2016}, and the last band was seen on (96) Aegle (Fig. \ref{Fig03}).}
\label{Fig08}
\end{figure}

We address the surface characteristics of large T-type asteroids in relation to those of low-albedo objects first by comparing their LNIR spectra. Figure \ref{Fig08} displays the spectra of (96) Aegle, (308) Polyxo \citep{Takir2012}, Jupiter Trojans \citep{Brown2016}, (1) Ceres \citep{DeSanctis2018}, and comet 67P/Churyumov-Gerasimenko (67P/C-G) \citep{Poch2020}, the last three of which were digitised from the published figures. Since (96) Aegle and (308) Polyxo show great consistency in their spectral behaviours and cover a broader range of wavelengths among the large T-types (Figure \ref{Fig05}), we used their spectra as a representative of the T-types.

A spectroscopic study of 16 Jupiter Trojans finds $\sim$3.1-$\mu$m N--H stretch and $\sim$3.4-$\mu$m organic features for a subpopulation with less-red colour \citep{Brown2016}. The nucleus of comet 67P/C-G from the {\it Rosetta} mission also reveals a prominent N--H band at 3.1--3.2 $\mu$m \citep{Poch2020} and a broad C--H absorption complex of organics between 3.3 and 3.6 $\mu$m \citep{Quirico2016}. The corresponding features may exist on our T-type (96) Aegle spectrum but to a lesser degree at $\sim$3.1 $\mu$m and $\sim$3.5 $\mu$m (Figs. \ref{Fig03} and \ref{Fig08})). This is quite surprising since such 3.1-$\mu$m N--H bands require the availability of NH$_{\rm 3}$ and/or N$_{\rm 2}$ ices in the accretion stage \citep{Lodders2003}. That is, the presence of the volatile bands requires parent bodies (or their constituents) to be originated from the outer solar system ($\gtrsim$5--10 au; \citealt{Lodders2004,Kurokawa2020}), far off the current positions of large T-type asteroids in the mid-asteroidal belt. The spectrum of C-type (1) Ceres from the {\it Dawn} mission contains a well-developed 2.7-$\mu$m band of Mg-rich phyllosilicates as large T-types due to the extensive aqueous alteration \citep{McSween2018}. However, its 3.1-$\mu$m band of NH$_{\rm 4}$ salts and organics and carbonate bands at 3.3--3.6 $\mu$m and 3.9--4.0 $\mu$m \citep{DeSanctis2018} implies that the early materials of (1) Ceres probably contained abundant CO$_{\rm 2}$/CO and NH$_{\rm 3}$ ices \citep{Carrozzo2018,Kurokawa2020} and again signals its connection to the outer region ($\gtrsim$10 au; \citealt{Fujiya2019}). 

T-types' LNIR continuum slope ($\sim$0.3 $\mu$m$^{\rm -1}$) is shallower than Jupiter Trojans ($\sim$0.9 $\mu$m$^{\rm -1}$) but comparable to the nucleus of 67P/C-G ($\sim$0.2 $\mu$m$^{\rm -1}$) and (1) Ceres ($\sim$0.3 $\mu$m$^{\rm -1}$). This similarity in slope may have implications for the close affinity of anhydrous (i.e., early) components of T-type asteroids to those in the primitive Ceres and comet 67P/C-G, which is again unfeasible if large T-type asteroids were formed and have been stored at their current warmer locations.
A OH-absorption band present on the T-types but not on the other low-albedo bodies (Fig. \ref{Fig08}) suggests that the T-type asteroids had abundant water ice at their birth and experienced higher-temperatures than the others for melting ice for some time \citep{McSween2002}.
It is unclear whether the lack of C--O bands on the T-type spectra is due to their formation distance being closer  than the CO$_{\rm 2}$ condensation front, the outgassing during the aqueous alteration, or the current orbital positions of the asteroids in the mid-asteroidal belt sublimating CO$_{\rm 2}$ ices from the near-surface layers \citep{Kurokawa2020}. However, given the similarities observed in the N--H and C--H bands and the background refractory materials, it is most likely that the large T-type asteroids originated outside of the main asteroid belt.

The 0.7-$\mu$m band is attributable to (but not definitive for) hydrated minerals harbouring iron oxides in phyllosilicates \citep{Vilas1989,Vilas1994}, which, together with much lower thermal background levels than in the LNIR, makes the 0.7-$\mu$m feature a tool to assess the aqueous alteration history of asteroids (\citealt{Rivkin2015b} and references therein). \citet{Rivkin2002} found that the observed frequencies of the 0.7-$\mu$m and 3.0-$\mu$m bands for low-albedo asteroids (the C, B, G, F, and P classes) seem proportional. \citet{Rivkin2012} report from the distribution of the 0.7 $\mu$m band that roughly 60 \% of the C-complex asteroids in the Sloan Digital Sky Survey Moving Object Catalogue might have experienced hydration. \citet{Fornasier2014} suggest that the difference in the detection rate of the 0.7-$\mu$m band for different taxonomic types likely reflects a range of their primitiveness. However, asteroids with sufficient signals of OH-absorption bands do not always accompany the 0.7-$\mu$m band \citep{Fornasier2014,Rivkin2015b,Usui2019}, which is often explained on account of the weak 0.7-$\mu$m band intensity (typically $\sim$1--6 \%; \citealt{Vilas1989,Fornasier2014}) and tends to disappear more rapidly than the 3.0-$\mu$m band with heating in laboratory experiments \citep{Hiroi1993,Hiroi1996}.

\begin{figure}[!t]
\centering
\includegraphics[width=9cm]{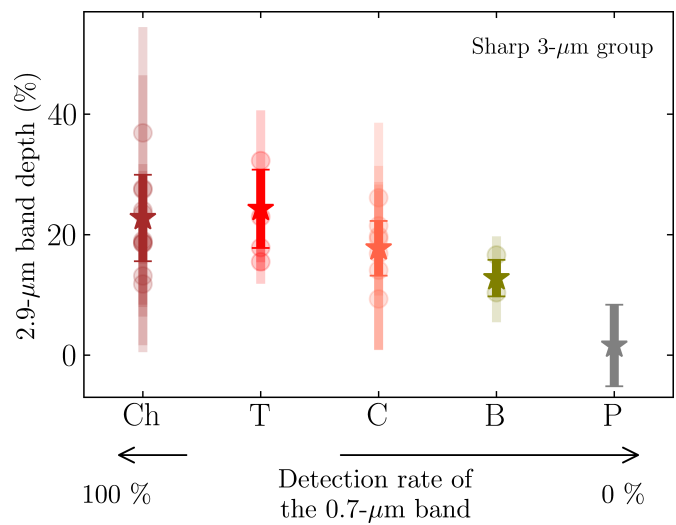}
\vspace{-.4in}
\caption{Band depth measured at 2.9 $\mu$m ($BD_{\rm 2.9}$) for asteroids in the sharp 3-$\mu$m group \citep{Takir2012}. Weighted average of each group is marked as a star symbol with 1$\sigma$ error. All calculation results and references are tabulated in Table \ref{ta2}. The detection rate of the 0.7-$\mu$m band was quoted from \citet{Fornasier2014}. $BD_{\rm 2.9}$ was not considered for T-type asteroids in the previous study.}
\label{Fig09}
\end{figure}

We examined a correlation between the 3.0-$\mu$m and 0.7-$\mu$m band depths for low-albedo asteroids, considering only large ($>$80 km in diameter) asteroids belonging to the sharp 3-$\mu$m group whose 3.0-$\mu$m band centre ($<$2.9 $\mu$m) indicates the phyllosilicate-dominated mineralogy \citep{Takir2012,Takir2015}. This limited sample selection would mitigate the influence of the water ice feature (at $\sim$3.1 $\mu$m) dominating the spectral shapes of other 3-$\mu$m groups, thus allowing us to trace the history of aqueous alteration on asteroids more reliably. The band depth at 2.9 $\mu$m was used to approximate the OH-band intensity for consistency with previous studies \citep{Takir2012,Takir2015} and calculated from  
\begin{equation}
BD_{\rm 2.9} = \frac{R_{\rm cont} - R_{\rm 2.9}}{R_{\rm cont}}~,
\label{eq:eq5}
\end{equation}
\noindent where $R_{\rm cont}$ and $R_{\rm 2.9}$ are the reflectances of the continuum and at 2.9 $\mu$m, respectively. $R_{\rm cont}$ was obtained by extending the linear regression line of the $K$-band (1.8--2.5 $\mu$m) to 2.9 $\mu$m. We measured $BD_{\rm 2.9}$ of (96) Aegle, (570) Kythera and (773) Irmintraud in Figure \ref{Fig05} and adopted the values of other asteroids from \citet{Takir2012} and \citet{Takir2015}. $BD_{\rm 2.9}$ values of asteroids in the same taxon were weighted-averaged. All calculation results and references used for Figure \ref{Fig09} are tabulated in Table \ref{ta2}. 

Figure \ref{Fig09} arranges $BD_{\rm 2.9}$ of the sharp 3-$\mu$m group asteroids in descending order: Ch (strongest) $\sim$ T $\rightarrow$ C $\rightarrow$ B $\rightarrow$ P (weakest). The number of asteroids used for the analysis is 6 for Ch-type, 4 for T-type, 7 for C-type, 2 for B-type, and 1 for P-type. This taxonomic order of $BD_{\rm 2.9}$ is in parallel to the order established by the 0.7-$\mu$m band detection rate  \citep{Fornasier2014}, albeit the authors consider all 3-$\mu$m spectral groups of asteroids in a wider range of sizes (down to $\sim$50 km) and T-type asteroids were not considered separately. The 0.7-$\mu$m band seems always present in Tholen G-types (corresponding to Ch-types in the Bus-DeMeo taxonomy) with the most prominent 3.0-$\mu$m band and becomes rarer from T-, C-, B-, finally to P-types (0 \%) with a nearly zero OH-band intensity. Figure \ref{Fig09} might thus support the association of the 0.7-$\mu$m band with the aqueous alteration but also link the prevalence of the 0.7-$\mu$m band to the degree of alteration. In this relationship, T-type asteroids on average might have experienced more prevalent aqueous alteration than most of the hydrous C-complex asteroids but to a roughly comparable extent to Ch-type asteroids. However, this relation should be regarded with some caution until a more significant number of asteroids are accumulated for all taxonomic types to minimise observational biases.  

\subsection{Surface texture of large T-type asteroids \label{sec:dis2}}

The environments asteroids have been exposed to shape their present-day compositional and physical surface properties \citep{Clark2002}. Once scattered by the surface, sunlights become partially linearly polarised. The resultant degree of linear polarisation ($P$) draws a bell-shaped dependence on the phase angle ($\alpha$, angle of Sun--asteroid--observer), often fitted by an empirical trigonometric function (\citealt{Cellino2015} for a review). This phase curve can be characterised by six parameters, among which those depicting the curve around the backscattering region (i.e., small $\alpha$) are well known to be sensitive to the texture and optical properties of the surfaces \citep{Wolff1975,Dollfus1975,Geake1986,Geake1990,Dollfus1998}.

\begin{figure}[!b]
\centering
\includegraphics[width=8.5cm]{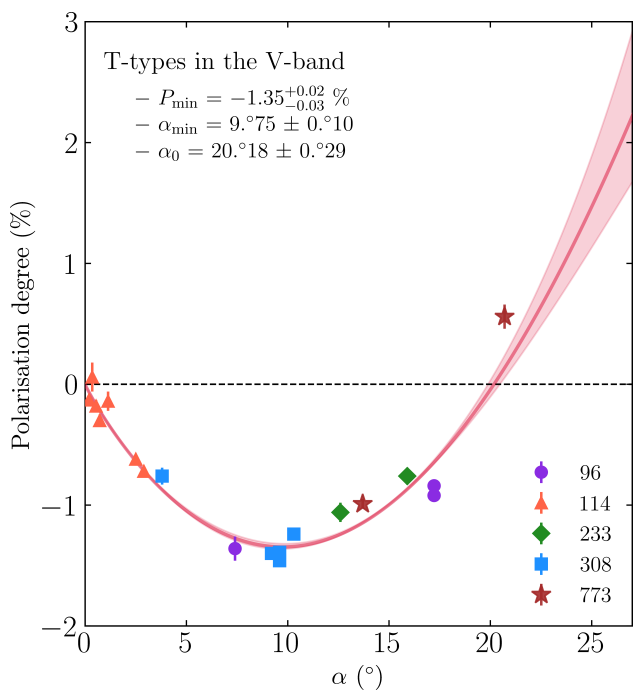}
\caption{Polarisation of large T-types' surfaces versus phase angle in the $V$ band. All data were quoted from the NASA/PDS APD \citep{Lupishko2022}. The red solid curve and shaded area indicate the average dependence curve derived from Eq. \ref{eq:eq6} and its 1$\sigma$ uncertainty. 
}
\label{Fig10}
\end{figure}

From the Asteroid Polarimetric Data (APD) provided by the NASA Planetary Data System (PDS) Small Bodies Node \citep{Lupishko2022}, we adopted $V$-band (the effective wavelength of 0.55 $\mu$m) data because of the largest number of observations therein, where five large T-type asteroids -- (96) Aegle, (114) Kassandra, (233) Asterope, (308) Polyxo, and (773) Irmintraud -- are available. The polarimetric phase curve, $P(\alpha)$, of the selected data was then obtained using the trigonometric function of \citet{Lumme1993}:
\begin{equation}
 P(\alpha) = b~\sin^{c_{\rm 1}}(\alpha) \cos^{c_{\rm 2}} \bigg(\frac{\alpha}{2}\bigg)  \sin(\alpha - \alpha_{\rm 0})~,
\label{eq:eq6}
\end{equation}
\noindent where $b$, $c_{\rm 1}$, $c_{\rm 2}$, and $\alpha_{\rm 0}$ are free parameters shaping the curve. $\alpha_{\rm 0}$ is the inversion angle where $P$ equals 0 \% and thus has different signs on its either side, that is, $P$ is negative at $\alpha$ < $\alpha_{\rm 0}$ and positive at $\alpha$ > $\alpha_{\rm 0}$. Figure \ref{Fig10} shows the average trend of the T-type asteroids. The fitting parameters are $b$ = 0.39 $\pm$ 0.03, $c_{\rm 1}$ = 0.93 $\pm$ 0.03, $c_{\rm 2}$ = (2.47 $\pm$ 0.10) $\times$ 10$^{\rm -11}$, and $\alpha_{\rm 0}$ = 20\fdg18 $\pm$ 0\fdg29. The minimum polarisation degree ($P_{\rm min}$) and its phase angle ($\alpha_{\rm min}$) were retrieved by differentiating the above equation and leaving it equal to zero: $P_{\rm min}$ = $-$1.35$^{\rm +0.02}_{\rm -0.03}$ \% and $\alpha_{\rm min}$ = 9\fdg75$^{\rm +0\fdg10}_{\rm -0\fdg10}$. 

\begin{figure}[!t]
\centering
\includegraphics[width=9cm]{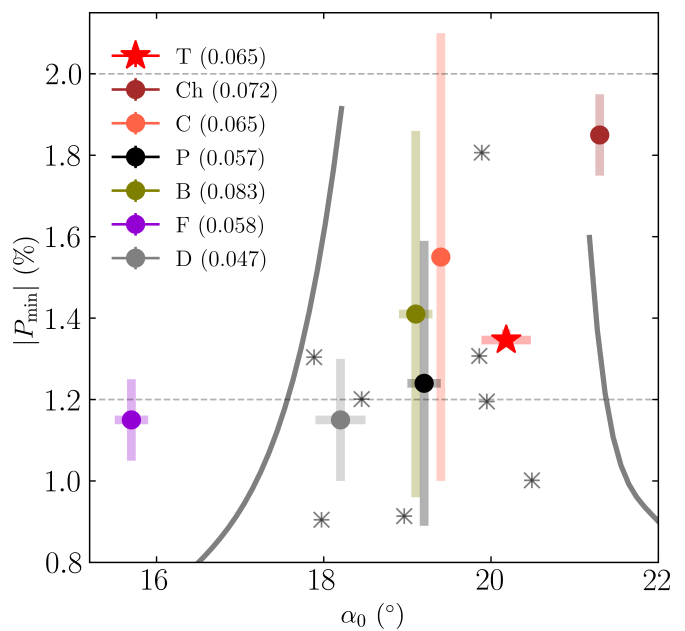}
\caption{Polarimetric parameter plot showing a relation between the absolute value of the minimum polarisation degree $|P_{\rm min}|$ and the inversion angle $\alpha_{\rm 0}$. Input parameters for other low-albedo asteroids were quoted from Table 3 in \citet{Belskaya2017}. The upper ($|P_{\rm min}|$ = 2.0 \%) and lower ($|P_{\rm min}|$ = 1.2 \%) dashed lines encase the $|P_{\rm min}|$ range of carbonaceous chondrites \citep{Zellner1977}. Thick solid curves indicate the boundaries of rocky--particulate surfaces (left curve) and particulate--fine-grained surfaces (right curve) derived from \citet{Geake1986}. Asterisks are the data for pulverised rocks and meteorites whose albedo is lower than 0.10 \citep{Geake1986}. Parentheses in the legend show the average geometric albedo of asteroids used here.
}
\label{Fig11}
\end{figure}

The retrieved polarimetric parameters of the T-type asteroids were then compared with those of low-albedo asteroids in different taxonomic types (Table 3 in \citealt{Belskaya2017}) in the plot of the absolute value of $P_{\rm min}$ ($|P_{\rm min}|$) versus $\alpha_{\rm 0}$ (Fig. \ref{Fig11}). Two dashed lines (2.0 and 1.2 \%) encase the $|P_{\rm min}|$ range of carbonaceous chondrites measured by \citet{Zellner1977}. Asterisks are the measurements of dark (albedo $<$ 0.10) pulverised rocks and meteorites \citep{Geake1986}. Thick solid curves partition the territories of rocky--particulate surfaces (left curve) and particulate--fine-grained surfaces (right curve) derived from the measurements of lunar regoliths \citep{Geake1986}. Numbers in parentheses inform the average geometric albedos of the asteroids used here. We found no one-to-one relation between the albedo and polarimetric parameters, as \citet{Zellner1976} reported that the polarimetric--albedo relationship saturates for very dark (albedo $\lesssim$ 0.06) asteroids. Within error limits, $|P_{\rm min}|$ values of all taxonomic types broadly overlap the range of carbonaceous chondrites, though the three darkest (F/D/P) types are located more on the borderline of the lower limit. 

Among the two parameters, the inversion angle $\alpha_{\rm 0}$ is more sensitive to the surface texture (i.e., the existence of sub-wavelength small grains on the surface; \citealt{Geake1990}). As such, the fact that the polarimetric parameters of large T-types are well in the region of dusty rocks can be interpreted as their surfaces being covered by particulate media, not by rocks or large fragments or fine-grained fluffy dust \citep{Geake1986}. This is well aligned with a general tendency of asteroidal surfaces: km-sized bodies as rubble piles of collisional fragments are covered by coarse surface grains with typical sizes of mm--cm range, whereas larger objects likely possess fine on-surface grains with sizes an order of 10--100 $\mu$m \citep{Gundlach2013,Vernazza2016,Beck2021}. 
The T-, Ch-, C-, P-, and B-type asteroids in the particulate surface region have more or less similar diameters of $\sim$120--170 km\footnote{All size information was quoted from the NASA/JPL Small Body Query Database (\url{https://ssd.jpl.nasa.gov/tools/sbdb\_query.html})} (where abnormally large (1) Ceres of 939.4 km, (2) Pallas of 513 km, and (10) Hygiea of 407.12 km in diameter were excluded from the count). The average size of the D-types used here is about half ($\sim$68 km), which might result in their observed $\alpha_{\rm 0}$ position near the coarser-grained territory. However, such a trend seems not applicable to the point of F-types located in the rocky surface region, albeit their average size of $\sim$130 km. It is uncertain for the moment whether surficial processes vastly different from most low-albedo asteroids have prevailed on the F-types or the interpretation of $\alpha_{\rm 0}$ is not a simple function of the grain size, especially for the dark asteroids of interest \citep{Belskaya2017}. Therefore, we only take the most straightforward message from Figure \ref{Fig11} that a similar location of the large T-types with low-albedo dusty rocks, meteorites, and other large-diameter asteroids in the polarimetric parameter plot would inform their particulate surfaces covered by $\sim$10--100 $\mu$m grains \citep{Gundlach2013}. 

\begin{table*}[!t]
\centering
\caption{Information on the meteorite spectra used in Figure \ref{Fig12}}
\vskip-1ex
\begin{tabular}{c|c|c|c|c|c|c}
\toprule
\hline
Panel & \multirow{2}{*}{Spectrum ID} & \multirow{2}{*}{Type} & \multirow{2}{*}{$\lambda_{\rm min}$$^{\rm (1)}$} & \multirow{2}{*}{$\lambda_{\rm max}$$^{\rm (2)}$} & \multirow{2}{*}{$a$$^{\rm (3)}$} & \multirow{2}{*}{Reference}\\
in Figure \ref{Fig12} & & & &  & & \\
\hline
\hline
\multirow{7}{*}{a} & bmr1mt186 & Cold Bokkeveld (CM2) & 1.4 & 25.1 &  \multirow{7}{*}{<35} & \multirow{7}{*}{RELAB$^{\rm (4)}$} \\
& bmr1mt187 & Nogoya (CM2) & 1.4 & 25.1& \\
& bmr1mt188 & Mighei (CM2) & 1.4 & 25.1 & \\
& bmr1mt190 & Murchison (CM2)& 1.4 & 25.1 & \\
& bir1mt218 & MET 01070 (CM1) & 0.8 & 28.6 & \\
& bir1mt220 & QUE 93005 (CM2) & 0.8 & 28.6 & \\
& bir1mt221 & Y-791198 (CM2) & 0.8 & 28.6 & \\
\hline
\multirow{4}{*}{b} & bir1mb064d1 & \multirow{4}{*}{Murchison (CM2)} & 0.8 & 99.7 & <25 & \multirow{4}{*}{RELAB} \\
& bir1mb064d2 & & 0.8 & 99.7 & 25--45 & \\
& bir1mb064d3 & & 0.8 & 99.7 & 45--75 & \\
& bir1mb064d4 & & 0.8 & 99.7 & 75--125 & \\
\hline
\multirow{7}{*}{c} & bir1mt025a & Tagish Lake (C2-ung) & 0.8 & 99.7 & <25 & \multirow{5}{*}{RELAB} \\
& bir1mt025b & Tagish Lake (C2-ung) & 0.8 & 99.7 & 25--125 & \\
& bir1mt264 & Alias (CI1) & 0.8 & 99.7 & <35 & \\
& bmr1mt191 & Orgueil (CI1) & 1.4 & 25.1 & <35 & \\
& bir2mx064p & Orgueil$+$Ivuna (CI1) & 0.8 & 25.1 & <125 & \\
   \cline{2-7}
  & 2478-7413$^{\rm (5)}$ & Ivuna (CI) & 0.3 & 25.1 & fine powders$^{\rm (6)}$ & \citet{Takir2013} \\
\hline
\bottomrule
\end{tabular}
\tablefoot{
\tablefoottext{\rm 1}{Starting wavelength for the spectrum in $\mu$m;}
\tablefoottext{\rm 2}{Ending wavelength for the spectrum in $\mu$m;} 
\tablefoottext{\rm 3}{the particle size in $\mu$m;} 
\tablefoottext{\rm 4}{RELAB database ({\url{http://www.planetary.brown.edu/relabdocs/relab.htm}});}
\tablefoottext{\rm 5}{Section number from the Smithsonian Institution, as described in \citet{Takir2013};}
\tablefoottext{\rm 6}{Ground, but not measured exactly.}
}
\label{t2}
\vskip-1ex
\end{table*}

\subsection{Comparison with carbonaceous chondrites \label{sec:dis3}}

Asteroids are expected to be a major source of meteorites \citep{Lipschutz1989}. Dark asteroids devoid of strong diagnostic absorption bands in the VNIR have been associated with carbonaceous chondrites (CCs), a group of meteorites relatively rich in C, H, N, and O \citep{Miyamoto1994,Burbine2002} that could have been relevant to habitability \citep{Kwok2016}. Among CCs, CI (Ivuna-type) and CM (Mighei-type) chondrites are of particular interest since they contain a prominent 3-$\mu$m OH absorption band, indicating aqueous alteration on their parent bodies \citep{Lebofsky1981,Beck2010,Cloutis2011a,Cloutis2011b,Takir2013,McAdam2015,Beck2021b}. It is hard to directly compare asteroids and meteorites primarily due to the possible difference in sampling regions (i.e., observed asteroidal surfaces susceptible to space weathering in contrast to meteorites originating from any parts of the parent bodies) and observing geometry \citep{Hiroi2001,Jedicke2004}. Nonetheless, connecting large T-type asteroids with CCs, particularly their LNIR reflectance spectra, sheds light on the composition of the bodies and the nature of thermal processes that occurred in the early solar system.

We set the spectrum of (96) Aegle as a representative of large T-type asteroids in our limited sample, as its spectral shape over the entire VNIR--LNIR range is almost identical to much-observed (308) Polyxo but with less noise in the spectrum (Fig. \ref{Fig08}). The whole spectrum was scaled to its visible geometric albedo (0.048; \citealt{Mainzer2019}) and compared with hydrated CM and CI chondrites. Since sample texture (e.g. powder and pellet) can significantly affect spectral slope and the depth of absorption features \citep{Ross1969,Johnson1973,Cloutis2018}, we selected meteorite spectra only in a particular/ground state in line with what T-type's polarimetric properties imply for their surface structure (Fig. \ref{Fig11}). Detailed information on the meteorite samples used for comparison is listed in Table \ref{t2}. 

\begin{figure}[!t]
\centering
\includegraphics[width=9cm]{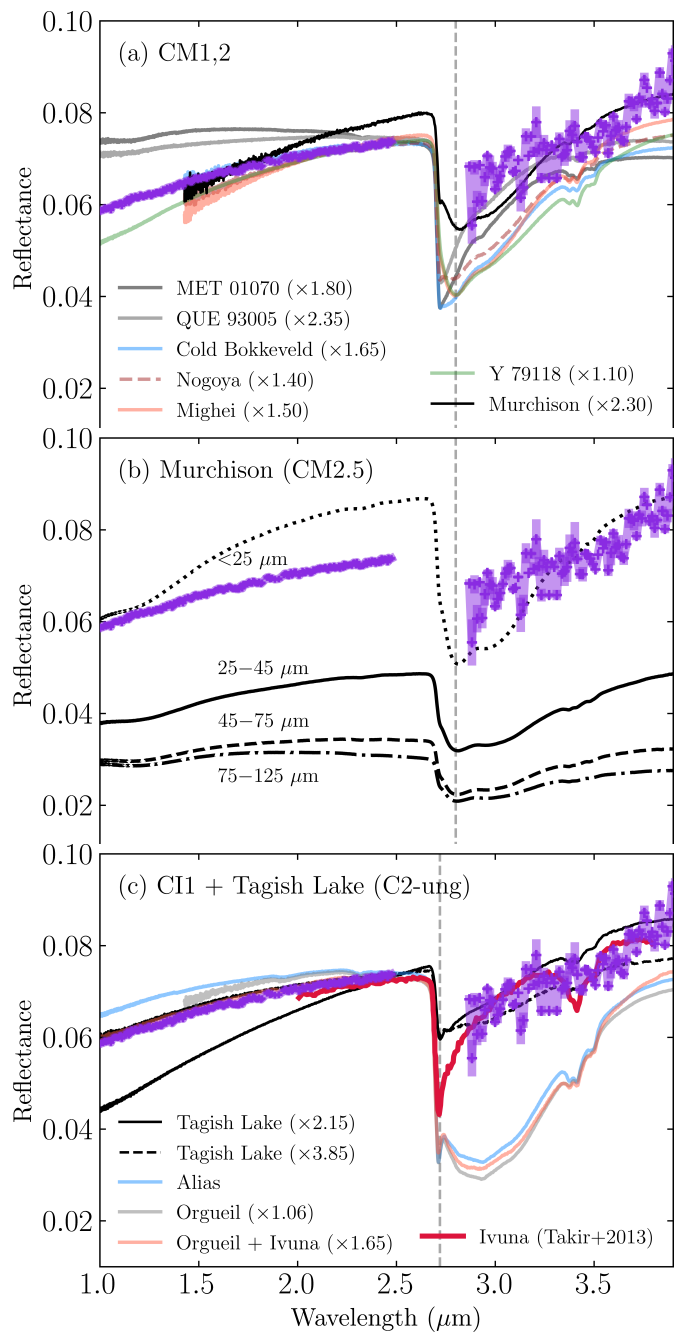}
\caption{Bidirectional reflectance of (a) CM chondrites in a range of hydration levels, (b) CM2 Murchison with different particle sizes and (c) CI1 and Tagish Lake chondrites, together with spectra of (96) Aegle from Figure \ref{Fig05}. Detailed information on the samples is listed in Table \ref{t2}. Vertical dashed lines inform absorption band centres at 2.83 $\mu$m for intermediate-to-least-hydrated CM chondrites and at 2.72 $\mu$m for CI and C2-ungrouped chondrites \citep{Takir2013}.
}
\label{Fig12}
\end{figure}

Figure \ref{Fig12}a plots the T-type spectrum with CM chondrites that account for the dominant number of hydrous CCs \citep{Zolensky1997} and have been nominated as a meteorite analogue for Ch-type asteroids \citep{Rivkin2015a,DeMeo2022}. Petrologic variations among CM chondrites correspond to a range of the intensity of aqueous alteration, based on which they can be classified into several subgroups (e.g. \citealt{Browning1996,Rubin2007}).  We selected CM samples at different hydration levels: MET 01070 $>$ QUE 93005 $>$ Cold Bokkeveld $>$ Nogoya $>$ Mighei $>$ Y-79118 $>$ Murchison in decreasing order, following the criteria of \citet{Rubin2007}. For consistency, the samples' particle size was unified as $<$35 $\mu$m. In Figure \ref{Fig12}a, strongly altered CM samples explain neither the continuum slope nor the absorption band shape of the T-type spectrum. Positing that spectra of dark asteroids often have bluer slopes than their meteorite analogues \citep{Beck2021}, likely due to the space weathering (change of the surface properties by solar-wind or micrometeoroid bombardments) on the asteroids' surfaces \citep{Lantz2017,Lantz2018}, CM samples showing far flatter slopes than the T-type would not be a good analogue candidate. Murchison (black curve) matches the overall T-type spectrum relatively better than the other CM samples. However, its reflectance seems too low ($\sim$2.3 times darker than the T-type) to be compensated by the difference in observing conditions (retrieved by \citealt{Beck2021a}) and has the OH bandwidth broader than the T-type spectra.

In Figure \ref{Fig12}b, CM2.5 Murchison samples in various particle sizes were compared with the T-type spectrum to see whether different sizes could diminish the observed discrepancies. The OH absorption feature looks most pronounced when the grain size is $<$25 $\mu$m but dampened for bigger grains, with the sample in 75--125 $\mu$m size showing the weakest band depth. As grain sizes increase, the sample spectrum becomes flattened and darker both in the VNIR and LNIR regions, as presented by \citet{Milliken2007} and \citet{Vernazza2016}. There is still a remaining mismatch in the spectral slope, overall brightness, and absorption band shape. Space weathering might alter band depth and centre (e.g. \citealt{Lantz2017}). However, 
realistic levels of irradiation for main-belt asteroids are not sufficient to change the band centres or shapes observably \citep{Hiroi2020, Prince2022}. CM chondrites would thus not be the best analogue for large T-type asteroids.

Finally, Figure \ref{Fig12}c illustrates the spectra of CI1 chondrites and Tagish Lake (C2-ungrouped). The abundance of phyllosilicates and fine-grained matrix in these CCs is generally much higher than in CM chondrites due to the advanced aqueous alteration on their parent bodies \citep{Wood2005,Cloutis2011a}. 
As alteration proceeds, phyllosilicates evolve from Fe-rich to Mg-rich; thereby the OH band centre moves from $\sim$2.83 $\mu$m to $\sim$2.72 $\mu$m, and its band shape becomes deeper, narrower, and more asymmetric  \citep{Takir2013}. Tagish Lake meteorites have been suggested as one of the best meteorites for D/T/P-type red asteroids based on their slope similarity \citep{Hiroi2001,DeMeo2022}. Our comparison shows that Tagish Lake spectra in small ($<$25 $\mu$m, the solid black line) and larger (25--125 $\mu$m , the dashed black line) particle sizes  are both much darker than the T-type spectrum. They also display either a much steeper VNIR slope (the smaller one) or shallower LNIR slope (the larger one) than the T-type's, though matching the OH-band shape better than CM chondrites. Three CI spectra from the RELAB database have VNIR slopes and overall reflectance consistent with the T-type spectrum (the smaller the grain size, the better match). We noted that their broad troughs around 3 $\mu$m denote the contamination of adsorbed terrestrial water molecules \citep{Gounelle2008,Beck2010,Takir2013} and instead referred to another fine-grained CI Ivuna spectrum from \citet{Takir2013}. Without requiring any adjustments, the new CI1 sample shows congruence with the T-type spectrum in the VNIR and LNIR spectral shapes, the OH-band shape (depth and width), and a subtle absorption between 3.3--3.6 $\mu$m. This meteorite has also been suggested as a meteorite analogue for T-type (308) Polyxo \citep{Takir2015}. 

Observations in mid-infrared wavelength (MIR; $\sim$5--38 $\mu$m) dispute such an association because  silicate emission features of D/T/P-like red asteroids resemble hydrated CO3 (Ornans-type) chondrites in small ($<$20 $\mu$m) sizes rather than Tagish Lake-like heavily hydrated CCs  \citep{Dotto2004,Vernazza2013}. However, at shorter wavelengths, particularly in the LNIR region relevant to the hydration process, such hydrated CO3 chondrites are unable to describe the OH absorption band centre and shape of the T-type spectrum\footnote{The RELAB Spectral Library interface provides visualisation of the meteorite spectra: CO3 ALH-77307 (\url{https://pds-speclib.rsl.wustl.edu/measurement.aspx?lid=urn:nasa:pds:relab:data\_reflectance:bnr1mp102}) and CO3 Y-82050 (\url{https://pds-speclib.rsl.wustl.edu/measurement.aspx?lid=urn:nasa:pds:relab:data\_reflectance:bnr1mp103})}. This could imply that, on the one hand, the aqueous alteration processes T-types and CO3's parent body(ies) encountered might have different natures \citep{Howard2009}. On the other hand, this discrepancy would emerge if two wavelengths sense different surface characteristics. Given that surfaces of D/T/P-type asteroids display comet-like silicate emission features \citep{Dotto2004,Emery2006}, the observed emissivity in the MIR could be more sensitive to the physical properties of small grains on the surfaces, such as the grain size and porosity \citep{Kwon2021,Kimura2022}. In comparison, the reflectance around the OH-absorption band might be more determined by the mineralogy of hydrous silicates \citep{Takir2013,Prince2022}. 

To summarise, CI1 chondrites appear to best match our representative spectra for large T-type asteroids in terms of the overall VNIR--LNIR spectral shape, reflectance, and absorption band properties. Samples of smaller particles of $<$35 $\mu$m (but not smaller than 25 $\mu$m) regardless of chondritic types always better explain the T-type spectrum, whereas coarser samples of $>$45 $\mu$m have too low reflectance and shallow spectral slopes than the asteroid data. Among the meteorite spectra compared here, the reflectance spectrum of large T-type asteroids appears most similar to CI chondrites with grain sizes of $\sim$25--35 $\mu$m, seemingly consistent with the implication of surface condition derived from the polarisation parameters. 

\subsection{Implications for the origin of T-type asteroids \label{sec:dis4}}

The large T-type asteroids probed in this study have red, featureless spectra similar to D- and P-type asteroids in the VNIR region and $L$-band continuum slopes comparable to (1) Ceres and comet 67P/C-G. This spectral context of T-type asteroids proposes their stronger affinity to red, dark small bodies originating from the outer part of the solar nebula than colocating main-belt asteroids. 

The apparent concordance with CI chondrites (Fig. \ref{Fig12}), along with the possible detection of the 3.1-$\mu$m N--H band on (96) Aegle (Fig. \ref{Fig08}) that requires NH$_{\rm 3}$ and/or N$_{\rm 2}$ ices at the formation epoch \citep{Lodders2003}, favours the inward migration of large T-types into the current orbits. CI chondrites are exceptional in almost every aspect: they retain the highest abundances of volatile elements (materials made of C, H, O, and N; \citealt{Bland2005,Vollstaedt2020}), deuterium in organics \citep{Wood2005,Alexander2007,Gourier2008,Remusat2010}, and water-to-rock ratios \citep{Brearley2006,Alexander2019,McDonough2021} relative to any other CCs. All the chemical information points out that CCs but CI chondrites originated from a narrow radial distance from the Sun (2.7--3.4 au, thus experiencing similar peak temperatures), whereas CI-like materials migrated from farther distances in the protoplanetary disk \citep{Alexander2007,Vollstaedt2020}. During aqueous alteration on the parent bodies, the formation of saponite over serpentine was favoured by higher water-to-rock ratios \citep{Zolensky1989,Kurokawa2020}. Saponite group phyllosilicates naturally became enriched in CI chondrites instead of serpentine groups dominating in CM chondrites \citep{Brearley2006}. These two products of aqueous alteration have a different band centre in the 0.7-$\mu$m region: serpentines at $\gtrsim$0.7 $\mu$m and saponites at 0.6--0.65 $\mu$m  \citep{Cloutis2011a,Cloutis2011b}. Regarding that the 0.7-$\mu$m serpentine band of CM chondrites well conforms to the spectra of hydrous (e.g. Ch-, C-, and B-types) main-belt asteroids \citep{Rivkin2015a,DeMeo2022}, the CI-like large T-types' band centres at 0.6--0.65 $\mu$m (Fig. \ref{Fig06}) may strengthen the idea that they would have undergone different alteration processes from that operating at main-belt distances. 

\begin{figure}[!b]
\centering
\includegraphics[trim=0 0 30 0, width=10.3cm]{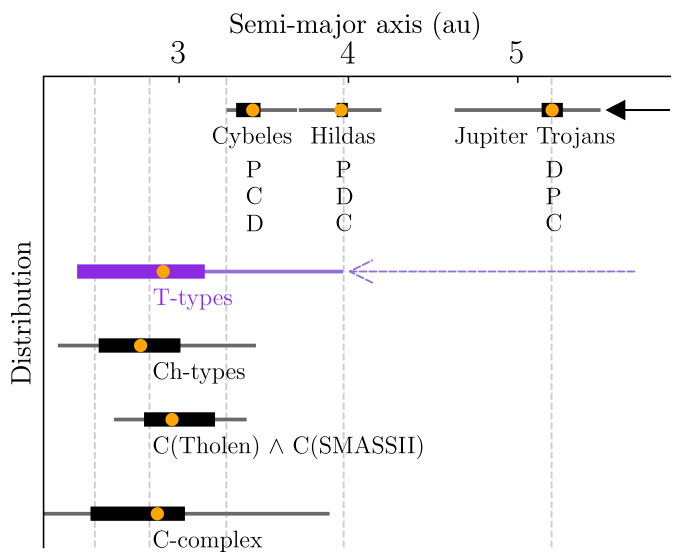}
\caption{Orbital distribution of low-albedo asteroids in different taxonomic types and dynamical groups  throughout the main asteroid belt and out to the orbit of Jupiter Trojans. Solid horizontal lines span the range of the minimum and maximum semi-major axes of the asteroids. Boxes delimit the standard deviation of the distribution of semi-major axes, and orange circles mark the average value. Spectral types constituting the Cybeles, Hildas, and Jupiter Trojans are given from top to bottom in decreasing order of dominance \citep{DeMeo2014}. Five vertical lines indicate semi-major axes of 2.502 au (3:1J resonance), 2.825 au (5:2J resonance), 3.279 au (2:1J resonance), 3.972 (3:2J resonance), and 5.2 au at Jupiter.}
\label{Fig13}
\end{figure}

If we suppose that large T-type asteroids were initially composed of D-type-like redder, darker materials (the most abundant taxonomic types beyond the main asteroid belt; \citealt{DeMeo2014}), aqueous alteration and enhanced space weathering on their way to the current orbital positions would entail a systematic change in the physical and compositional properties of the observed surfaces. The formation of opaque minerals (e.g. magnetite) as products of aqueous alteration would flatten the continuum slope \citep{Loeffler2002}. An extensive aqueous alteration may also cause grain growth \citep{Jones2006}, and for low-reflectance materials, the spectral slope becomes flattened as particle size increases \citep{Milliken2007,Vernazza2016}. The plausible size range of the regolith on top of large T-types constrained in this study ($\sim$25--35 $\mu$m) is larger than that of D-type Jupiter Trojans, for which previous studies retrieved that their spectra seem explained by fine grains of $<$25 $\mu$m \citep{Emery2004,Emery2006}. The larger on-surface grains of T-type asteroids might also demonstrate a similar band shape but weaker 10-$\mu$m silicate intensity of T-type (308) Polyxo than D-type (624) Hektor \citep{Dotto2004,Emery2006}. It seems that space weathering in the inner solar system is also capable of turning D-type's red slopes to C- and P-like shallower slopes \citep{Vernazza2013,Lantz2017,Lantz2018,Gartrelle2021}. \citet{Hasegawa2021} recently report a change in (596) Scheila’s continuum slope since its impact in 2010 from the T-type before the event to the D-type-like redder slope and therefrom suggest that the D-type slope would represent a less-evolved sub-surface layer of the weathered T-type asteroid. The weak correlation of the slope redness favourable for smaller T-types at a closer distance to the Sun (i.e., asteroids likely having experienced collisions in more recent times; \citealt{Bottke2005}) may also support this D--T connection. 

Consequently, we propose that the current location of large T-type asteroids is not representative of where they originally formed. Figure \ref{Fig13} illustrates the current orbital distribution of dark, red asteroids, with arrows showing the possible emplacements. The inward arrow pointing to Jupiter Trojans indicates the possibility that some of them were captured Trans-Neptunian Objects (\citealt{Levison2020} and references therein). The dominant spectral types of each group are listed in decreasing order from top to bottom \citep{DeMeo2014}. D-types, which tend to dominate in the outer part of the main belt and beyond, are generally thought to have a spectral appearance similar to that of comet nuclei (e.g. \citealt{DeMeo2008,Poch2020,Kareta2021}).
In the absence of C--O bands for the large T-types, we are unable to constrain their original location relative to the CO and CO$_{\rm2}$ condensation front ($\gtrsim$10 au;  \citealt{Fayolle2011,Fujiya2019}) solely based on their reflectance spectra because the observed dearth of volatiles could be either intrinsic (due to their formation distance at $<$10 au) or extrinsic (thermal conditions in their present orbits too hot for such primitive materials to survive; \citealt{Kurokawa2020}). We thus inferred their place of origin from the high spectral compatibility of T-types with CI chondrites.
It has been suggested that CI-like materials most likely originated outside of a pressure trap near the proto-Jupiter at $\sim$3--5 au in the solar nebula \citep{vanKooten2019,Larsen2020,Brasser2020,vanKooten2021}.
CI chondrites typically contain lower volatile contents \citep{Vollstaedt2020} and higher water-to-rock ratios \citep{Wood2005,Brearley2006} than most comets. However, recent dynamical studies have shown that comets could form over a wide range of semi-major axes in their early phases \citep{Brasser2013}. 
There are also differences in the deuterium-to-hydrogen (D/H) ratios of comets in water ranging from low ratios as terrestrial values to several times higher values (\citealt{Lis2019} and references therein), where the D/H ratios of CI chondrites are consistent with those of comets with terrestrial ratios  \citep{Gounelle2008,Alexander2017}. Given that the D/H ratio is a proxy for the temperature at the formation site \citep{Meier1999} and tends to increase with increasing distance from the Sun \citep{Drouart1999}, such consistency implies that the formation region of CI chondrites could overlap with the warmer (i.e., inner) part of comet formation range.

Taken as a whole, we tentatively conclude that large T-type asteroids may have been dislodged from the outside of the proto-Jupiter ($\sim$3--5 au) but not so far as the cryogenic Kuiper Belt ($\gtrsim$35 au), where D-type Jupiter Trojans \citep{Brown2016} and parent bodies of short-period comets likely formed \citep{Dones2015}. Hence, we propose that large T-type asteroids might originate somewhere between these two regions, near the very rough central point of comet formation distance ($\sim$10 au; \citealt{A'Hearn2012}). 
\\

\section{Summary \label{sec:sum}}

We presented new $L$-band spectroscopic observations of two large T-type asteroids, (96) Aegle and (570) Kythera, from the Subaru telescope in September 2020. Combining reflectance spectra of other T-type asteroids publicly available, we examined the surface characteristics to constrain their place of origin. The main results of the analysis are as follows.

\begin{enumerate}

\item Our targets exhibit red $L$-band slopes: (0.30$\pm$0.04) $\mu$m$^{\rm -1}$ for (96) Aegle and (0.31$\pm$0.03) $\mu$m$^{\rm -1}$ for (570) Kythera. The contribution of thermal excess to the observed reflectance was negligible for both targets, except at the long ends of the $L$-band region ($\gtrsim$3.8 $\mu$m). We found a discontinuity of the spectral curvature between the $K$-band (concaving down) and $L$-band (linear increase) for both asteroids, indicating the presence of an OH absorption band. We also found a possible N--H stretch band at $\sim$3.14 $\mu$m and C--H band of organic materials over 3.3--3.6 $\mu$m for (96) Aegle, but no further absorption features stronger than the $\pm$5 \% calibration uncertainty.
\\

\item We compared the targets' VNIR--LNIR reflectance spectra with three other large T-types whose $L$-band reflectance was measured: (308) Polyxo, (596) Scheila, and (773) Irmintraud. An OH-absorption band appears on their spectra but (596) Scheila. The linear increase of the reflectance over 2.9--3.2 $\mu$m makes the large T-types belong to the sharp 3-$\mu$m group \citep{Takir2012} whose surface is dominated by phyllosilicates. (96) Aegle and (308) Polyxo possess a nearly identical spectral shape across the VNIR and LNIR regions. The slope of (596) Scheila has changed from T- to D-type since its 2010 impact event, but the OH-band signal is missing. \\

\item Four out of eight large T-types appear to contain an absorption feature near 0.6--0.65 $\mu$m likely associated with iron oxides in phyllosilicates. For 21 T-type asteroids in all sizes (except Jupiter Trojans), we found that T-type NEOs tend to have the reddest VNIR slope among the asteroids considered, followed by intermediate-sized and larger ($>$80 km in diameter) asteroids for the shallowest. 
We infer that the phase reddening effect of dark asteroid surfaces may not be responsible for this colour distribution because the asteroids' phase angles at the observing epochs and their slopes are apparently uncorrelated (Spearman's correlation $r$ = $-$0.029). T-type NEOs inside 2 au possess redder slopes than their counterparts with main-belt semi-major axes. Other than these, no trends are seen as statistically significant within the current observations of T-type asteroids. \\

\item The large T-types retain an overall LNIR slope comparable to (1) Ceres and comet 67P/Churyumov-Gerasimenko but less red than Jupiter Trojans. Except for the possible absorption features of (96) Aegle, the large T-type asteroids are overall devoid of diagnostic 3.1-$\mu$m N--H band of ammonium or 3.9-$\mu$m C--O band of carbonates otherwise seen on these dark small bodies. \\

\item For large asteroids belonging to the sharp 3-$\mu$m group, we found a positive relationship between the average 2.9-$\mu$m band depth of a taxonomic type and the detection rate of the 0.7-$\mu$m band therein, supporting the utility of the 0.7-$\mu$m band to diagnose the intensity of aqueous alteration. In this context, the T-type might have experienced aqueous alteration roughly comparable to the Ch-type but more extensively than most hydrated main-belt asteroids. 
\\

\item We derived the minimum polarisation degree ($P_{\rm min}$ = $-$1.35$^{+0.02}_{-0.03}$) and the inversion angle ($\alpha_{\rm 0}$ = 20\fdg18 $\pm$ 0\fdg29) of large T-type asteroids from their polarimetric dependence on phase angle. Their polarimetric parameters are in the ranges of dark (albedo $<$ 0.1) pulverised rocks and dusty carbonaceous meteorites, indicating the T-types' particular surface texture. 
\\

\item The reflectance spectra of large T-type asteroids show the best match with CI chondrites with grain sizes of $\sim$25--35 $\mu$m among CCs in terms of the overall spectral slope, band shape and centre, and reflectance. CM chondrites able to explain hydrous C-complex main-belt asteroids seem less consistent with those spectral characteristics than CI chondrites. Meteorite spectra in small grains always better describe the asteroid data than coarser grains. \\

\item Compiling all results relevant to T-type asteroids currently available, we tentatively conclude that the current orbital positions of large T-type asteroids would not reflect their place of origin and they might be dislodged from roughly around 10 au. 

\end{enumerate}

Substantial mixing in the early solar system might have left its mark in the transitional natures of T-type asteroids connecting D- and C-type asteroids. Our study likely adds additional observational evidence supporting the implantation from outward of the main asteroid belt \citep{Levison2009,DeMeo2014}. We expect that systematic spectroscopic observations of T-type asteroids in a wide range of sizes and orbital distances using the {\it James Webb Space Telescope} will provide intriguing insights into the structural evolution of the solar system.
\\

\begin{acknowledgements}

We are grateful to Yuhei Takagi for his assistance and expert help with the 2020 Subaru observing run. Y.G.K gratefully acknowledges the support of the Alexander von Humboldt Foundation. S.F. acknowledges the financial support from the France Agence Nationale de la Recherche (programme Classy, ANR-17-CE31-0004). M.I. was supported by the NRF funded by the Korean Government (MEST) grant No. 2018R1D1A1A09084105. J.A. acknowledges funding by the Volkswagen Foundation and funding from the European Union’s Horizon 2020 research and innovation program under grant agreement No 75390 CAstRA. This research is based on data collected at Subaru Telescope, which is operated by the National Astronomical Observatory of Japan. We are honored and grateful for the opportunity of observing the Universe from Maunakea, which has the cultural, historical and natural significance in Hawaii.
\end{acknowledgements}


\begin{appendix}
\section{Summary of the asteroids' properties and their references \label{sec:app1}}

\begin{table*}[!h]
\small
\centering
\caption{Profiles of T-type asteroids used in Figure \ref{Fig07} and references of the original spectra}
\begin{sideways}
\begin{tabular}{lcc|cc|ccc|cc|c}
\toprule
\hline
Ast. Name & Tholen & SMASSII & $D_{\rm e}$$^{\rm (1)}$ & $p_{\rm V}$$^{\rm (2)}$ & $a$$^{\rm (3)}$ & $e$$^{\rm (4)}$ & $i$$^{\rm (5)}$ & $R_{\rm 0.8}$/$R_{\rm 0.5}$ & $R_{\rm 3.2}$/$R_{\rm 2.9}$ & Reference\\
\hline
\hline
(96) Aegle & T & T & 177.7 & 0.048 & 3.049 & 0.141 & 15.982 & 1.203$\pm$0.001  & 1.092$\pm$0.010 & \citet{DeMeo2009}, this study \\
(114) Kassandra & T & Xk & 94.2 & 0.088 & 2.676 & 0.138 & 4.945 & 1.185$\pm$0.001 & $-$ & \citet{DeMeo2009} \\
(233) Asterope & T & K & 99.7 & 0.093 & 2.660 & 0.100 & 7.691 & 1.196$\pm$0.001 & $-$ & \citet{DeMeo2009} \\
(308) Polyxo & T & T & 128.578 & 0.046 & 2.749 & 0.040 & 4.362 & 1.244$\pm$0.001 & 1.161$\pm$0.028 & \citet{DeMeo2009}, \citet{Takir2012} \\
(570) Kythera & ST & T & 87.486 & 0.044 & 3.420 & 0.119 & 1.816 & 1.235$\pm$0.001 & 1.099$\pm$0.013 & \citet{DeMeo2009}, this study \\
(595) Polyxena & $-$ & T$^{(\dagger)}$ & 90.647 & 0.096 & 3.206 & 0.064 & 17.831 & 1.250$\pm$0.001 & $-$ & \citet{Lazzaro2004} \\
(596) Scheila & PCD & T & 159.726 & 0.037 & 2.929 & 0.163 & 14.658 & 1.191$\pm$0.001 & 1.063$\pm$0.003 & \citet{DeMeo2009}, \citet{Yang2011} \\
(773) Irmintraud & D & T & 91.672 & 0.048 & 2.859 & 0.079 & 16.667 & 1.225$\pm$0.001 & 1.028$\pm$0.004 & \citet{DeMeo2009}, \citet{Kanno2003} \\
(1471) Tornio & $-$ & T & 28.719 & 0.085 & 2.716 & 0.119 & 13.607 & 1.241$\pm$0.002 & $-$ & \citet{DeMeo2009} \\
(2271) Kiso & $-$ & T & 31.229 & 0.055 & 2.762 & 0.061 & 3.389 & 1.231$\pm$0.002 & $-$ & $'$$'$ \\
(2813) Zappala & $-$ & T & 32.040 & 0.062 & 3.136 & 0.154 & 14.760 & 1.279$\pm$0.002 & $-$ & $'$$'$ \\
(2857) NOT & $-$ & T & 9.313 & 0.169 & 2.401 & 0.095 & 5.733 & 1.223$\pm$0.001 & $-$ & $'$$'$ \\
(2929) Harris & $-$ & T & 16.176 & 0.155 & 3.120 & 0.067 & 14.891 & 1.230$\pm$0.002 & $-$ & $'$$'$ \\
(3254) Bus & $-$ & T & 31.104 & 0.073 & 3.970 & 0.155 & 4.412 & 1.243$\pm$0.001 & $-$ & $'$$'$ \\
(3311) Podobed & $-$ & T & 17.336 & 0.071 & 2.787 & 0.040 & 0.931 & 1.208$\pm$0.003 & $-$ & $'$$'$ \\
(3443) Leetsungdao & $-$ & T & 8.852 & 0.132 & 2.394 & 0.306 & 12.709 & 1.225$\pm$0.001 & $-$ & $'$$'$ \\
(85490) 1997 SE5$^{\rm (6)}$ & $-$ & T & $-$ & $-$ & 3.762 & 0.661 & 2.575 & 1.232$\pm$0.001 & $-$ & $'$$'$ \\
(138911) 2001 AE2$^{\rm (6)}$ & $-$ & T & $-$ & $-$ & 1.350 & 0.081 & 1.662 & 1.296$\pm$0.004 & $-$ & $'$$'$ \\
(162998) 2001 SK162$^{\rm (6)}$ & $-$ & T & 0.875 & 0.161 & 1.925 & 0.475 & 1.681 & 1.286$\pm$0.004 & $-$ & $'$$'$ \\
2001 UU92$^{\rm (6)}$ & $-$ & T & $-$ & $-$ & 3.174 & 0.666 & 5.369 & 1.231$\pm$0.006 & $-$ & $'$$'$ \\
2001 YE1$^{\rm (6)}$ & $-$ & T & $-$ & $-$ & 1.914 & 0.501 & 4.456 & 1.304$\pm$0.002 & $-$ & $'$$'$\\
\hline
\bottomrule
\label{ta1}
\end{tabular}
\end{sideways}
\tablefoot{
\tablefoottext{\rm 1}{Effective body diameter in km;}
\tablefoottext{\rm 2}{Geometric albedo;} 
\tablefoottext{\rm 3}{Semi-major axis in au;} 
\tablefoottext{\rm 4}{Eccentricity;}
\tablefoottext{\rm 5}{Inclination in degree;}
\tablefoottext{\rm 6}{Near-Earth Objects (NEOs);} 
\tablefoottext{\rm $\dagger$}{Initially not observed by \citet{Bus2002}, but classified as T type in this taxonomic system.} The physical and dynamical properties are quoted from the NASA/JPL Small-Body Database Lookup (\url{https://ssd.jpl.nasa.gov/tools/sbdb\_lookup.html#/} and WISE database \citep{Masiero2011}).
}
\end{table*}

\begin{table*}[!h]
\centering
\caption{Profiles of sharp 3-$\mu$m group asteroids used in Figure \ref{Fig09} and reference of the 2.9-$\mu$m band depth ($BD_{\rm 2.9}$)}
\vskip-1ex
\begin{tabular}{cc|cc|c}
\toprule
\hline
Taxonomy & Ast. Name & $D$$^{\rm (1)}$ & $BD_{\rm 2.9}$ & Measurement\\
\hline
\hline
\multirow{11}{*}{Ch} & (13) Egeria & 202.636 & 27.66$\pm$2.91 & \citet{Takir2015} \\
 & (41) Daphne & 205.495 & 18.58$\pm$10.59 & $'$$'$ \\
 & (98) Ianthe & 132.788 & 19.03$\pm$3.85 & $'$$'$ \\
 & (211) Isolda & 141.125 & 18.74$\pm$2.68 & $'$$'$ \\
 & (34) Circe & 132.992 & 13.15$\pm$6.75 & \citet{Takir2012} \\
 & (48) Doris & 216.473 & 23.50$\pm$3.65 & $'$$'$ \\
 & (91) Aegina & 103.402 & 27.48$\pm$26.99 & $'$$'$ \\
 & (104) Klymene & 136.553 & 11.82$\pm$10.18 & $'$$'$ \\
 & (121) Hermione & 209.000 & 24.02$\pm$7.7 & $'$$'$ \\
 & (130) Elektra & 180.652 & 36.90$\pm$9.57 & $'$$'$ \\
 & (187) Lamberta & 147.294 & 18.56$\pm$10.18 & $'$$'$ \\
 \cline{1-5}
\multirow{3}{*}{T} & (96) Aegle & 177.774 & 15.51$\pm$3.64 & this study \\
 & (308) Polyxo & 128.578 & 17.87$\pm$3.86 & \citet{Takir2015} \\
 & (570) Kythera & 87.486 & 32.293$\pm$8.35 & this study\\
 & (773) Irmintraud & 91.672 & 22.97$\pm$7.53 & this study \\
 \cline{1-5}
\multirow{7}{*}{C} & (36) Atalante & 132.842 & 21.49$\pm$6.82 & \citet{Takir2012} \\
 & (54) Alexandra & 160.120 & 26.13$\pm$5.28 & $'$$'$ \\
 & (120) Lachesis & 155.132 & 14.11$\pm$13.23 & $'$$'$ \\
 & (334) Chicago & 198.770 & 9.33$\pm$8.50 & $'$$'$ \\
 & (511) Davida & 270.327 & 19.32$\pm$9.42 & $'$$'$ \\
 & (1015) Christa & 82.350 & 19.75$\pm$18.84 & $'$$'$ \\
 & (488) Kreusa & 168.117 & 16.86$\pm$3.12 & \citet{Takir2015} \\
 \cline{1-5}
\multirow{2}{*}{B} & (2) Pallas & 513.000 & 16.61$\pm$3.10 & this study \\
 & (704) Interamnia & 306.313 & 10.37$\pm$4.91 & \citet{Takir2012} \\
 \cline{1-5}
P & (140) Siwa & 109.790 & 1.60$\pm$6.77 & \citet{Takir2012} \\
\hline
\bottomrule
\label{ta2}
\end{tabular}
\tablefoot{The $D$ values stand for the effective body diameter in km, which were quoted from the NASA/JPL Small-Body Database Lookup (\url{https://ssd.jpl.nasa.gov/tools/sbdb\_lookup.html#/}).
}
\vskip-1ex
\end{table*}

\end{appendix}


\begin{thebibliography}{}

\bibitem[A'Hearn et al.(2012)]{A'Hearn2012} A'Hearn, M. F., Feaga, L. M., Keller, H. U., et al. 2012, \apj, 758, 29
\bibitem[Alexander et al.(2007)]{Alexander2007} Alexander, C. M. O'D., Fogel, M., Yabuta, H., \& Cody, G. D. 2007, \gca, 71, 4380
\bibitem[Alexander(2017)]{Alexander2017} Alexander, C. M. O'D. 2017, Phil. Trans. R. Soc. A, 375:20150384
\bibitem[Alexander(2019)]{Alexander2019} Alexander, C. M. O'D. 2019, \gca, 254, 277
\bibitem[Beck et al.(2010)]{Beck2010} Beck, P., Quirico, E., Montes-Hernandez, G., et al. 2010, \gca, 74, 4881
\bibitem[Beck et al.(2021a)]{Beck2021a} Beck, P., Schmitt, B., Potin, S., et al. 2021a, \icarus, 354, 114066
\bibitem[Beck et al.(2021b)]{Beck2021b} Beck, P., Eschrig, J., Potin, S., et al. 2021b, \icarus, 357, 114125
\bibitem[Beck \& Poch(2021)]{Beck2021} Beck, P., \& Poch, O. 2021, \icarus, 365, 114494
\bibitem[Belskaya et al.(2017)]{Belskaya2017} Belskaya, I. N., Fornasier, S., Tozzi, G. P. 2017, \icarus, 284, 30
\bibitem[Binzel et al.(2004)]{Binzel2004} Binzel, R. P., Rivkin, A. S., Stuart, J. S., et al. 2004, \icarus, 170, 259
\bibitem[Bland et al.(2005)]{Bland2005} Bland, P. A., Alard, O., Benedix, G. K., et al. 2005, PNAS, 102, 13755
\bibitem[Bottke et al.(2005)]{Bottke2005} Bottke, W. F., Durda, D. D., Nesvorn{\'y}, D., et al. 2005, \icarus, 175, 111
\bibitem[Bowell et al.(1989)]{Bowell1989} Bowell, E., Hapke, B., Domingue, D., et al. 1989, Asteroids II, 524
\bibitem[Brasser \& Morbidelli(2013)]{Brasser2013} Brasser, R., \& Morbidelli A. 2013, \icarus, 225, 40
\bibitem[Brasser \& Mojzsis(2020)]{Brasser2020} Brasser, R., \& Mojzsis, S. J. 2020, Nat. Ast., 4, 492
\bibitem[Brearley(1995)]{Brearley1995} Brearley, A. J. 1995, \gca, 59, 2291
\bibitem[Brearley(2006)]{Brearley2006} Brearley, A. J. 2006, In Meteorites and the early solar system II, Lauretta, D. S., \& McSween, H. Y. Jr., Eds., Tucson, Arizona: The University of Arizona Press, 584
\bibitem[Britt et al.(1992)]{Britt1992} Britt, D. T., Bell, J. F., Haack, H., et al. 1992, Meteoritics, 27, 207
\bibitem[Brown(2016)]{Brown2016} Brown, M. E. 2016, \aj, 152, 159
\bibitem[Browning et al.(1996)]{Browning1996} Browning, L., McSween H. Y., \& Zolensky, M. E. 1996, \gca, 60, 2621
\bibitem[Burbine et al.(2002)]{Burbine2002} Burbine, T. H., McCoy, T. J., Meibom, A., Gladman, B., \& Keil, K. 2002, Asteroids III, 653
\bibitem[Bus \& Binzel(2002)]{Bus2002} Bus, S. J., \& Binzel, R. P. 2002, \icarus, 158, 146
\bibitem[Carrozzo et al.(2018)]{Carrozzo2018} Carrozzo, F. G., De Sanctis, M. C., Raponi, A., et al. 2018, Sci. Adv., 4, e1701645
\bibitem[Cellino et al.(2015)]{Cellino2015} Cellino, A., Gil-Hutton, R., \& Belskaya, I. N. 2015, Polarimetry of Stars and Planetary Systems, eds. L. Kolokolova, J. Hough, \& A. Levasseur-Regourd (Cambridge: Cambridge University Press), 360
\bibitem[Charnoz \& Morbidelli(2007)]{Charnoz2007} Charnoz, S., \& Morbidelli, A. 2007, \icarus, 188, 468
\bibitem[Clark(1981)]{Clark1981} Clark, R. N., 1981, JGR, 86, 3074
\bibitem[Clark et al.(2002)]{Clark2002} Clark, R. N., Hapke, B., Pieters, C., et al. 2002, Asteroids III, 585
\bibitem[Clark et al.(2007)]{Clark2007} Clark, R. N., Swayze, G. A., Wise, R., et al. 2007. USGS digital spectral library,  231 (\url{http://speclab.cr.usgs.gov/spectral.lib06})
\bibitem[Cloutis et al.(2011a)]{Cloutis2011a} Cloutis, E. A., Hiroi, T., Gaffey, M. J., Alexander, C. M. O' D., \& Mann, P. 2011, \icarus, 212, 180
\bibitem[Cloutis et al.(2011b)]{Cloutis2011b} Cloutis, E. A., Hudon, P., Hiroi, T., Gaffey, M. J., \& Mann, P. 2011, \icarus, 216, 309
\bibitem[Cloutis et al.(2018)]{Cloutis2018} Cloutis, E. A., Pietrasz, V. B., Kiddell, C., et al. 2018, \icarus, 305, 203
\bibitem[Cruikshank \& Kerridge(1992)]{Cruikshank1992} Cruikshank, D.P., \& Kerridge, J. F. 1992, Organic Material: Asteroids, Meteorites, and Planetary Satellites, In Exobiology in Solar System Exploration, G. Carle, D. Schwartz, and J. Huntington, eds., NASA SP-512, 159-176 
\bibitem[Cruikshank et al.(2001)]{Cruikshank2001} Cruikshank, D. P., Dalle Ore, C. M., Roush, T. L., et al. 2001, \icarus, 153, 348
\bibitem[Delbo(2004)]{Delbo2004} Delbo, M. 2004, PhD thesis, Free University of Berlin
\bibitem[DeMeo \& Binzel(2008)]{DeMeo2008} DeMeo, F. E., \& Binzel, R. P. 2008, \icarus, 194, 436
\bibitem[DeMeo et al.(2009)]{DeMeo2009} DeMeo, F. E., Binzel, R. P., Slivan, S. M., \& Bus, S. J. 2009, \icarus, 202, 160
\bibitem[DeMeo \& Carry(2013)]{DeMeo2013} DeMeo, F. E., \& Carry, B. 2013, \icarus, 226, 723
\bibitem[DeMeo \& Carry(2014)]{DeMeo2014} DeMeo, F. E., \& Carry, B. 2014, \nat, 505, 629
\bibitem[DeMeo et al.(2022)]{DeMeo2022} DeMeo, F. E., Burt, B. J., Marsset, M., et al. 2022, \icarus, 380, 114971
\bibitem[De Luise et al(2010)]{DeLuise2010} De Luise, F., Dotto, E., Fornasier, S., et al. 2010, \icarus, 209, 586
\bibitem[De Sanctis et al.(2018)]{DeSanctis2018} De Sanctis, M. C., Ammannito, E., Carrozzo, F. G., et al. 2018, M\&PS, 53, 1844
\bibitem[Dollfus \& Geake(1975)]{Dollfus1975} Dollfus A., \& Geake J., 1975, LPSC, 3, 2749
\bibitem[Dollfus(1998)]{Dollfus1998} Dollfus, A. 1998, \icarus, 136, 69
\bibitem[Dones et al.(2015)]{Dones2015} Dones, L., Brasser, R., Kaib, N., \& Rickman, H. 2015, \ssr, 197, 191
\bibitem[Dotto et al.(2004)]{Dotto2004} Dotto, E., Barucci, M. A., Brucato, J. R., M{\"u}ller, T. G., \& Carvano, J. 2004, \aap, 427, 1081
\bibitem[Drouart et al.(1999)]{Drouart1999} Drouart, A., Dubrulle, B., Gautier, D., \& Robert, F. 1999, \icarus, 140, 129
\bibitem[Emery \& Brown(2003)]{Emery2003} Emery, J. P., \& Brown, R. H. 2003, \icarus, 164, 104
\bibitem[Emery \& Brown(2004)]{Emery2004} Emery, J. P., \& Brown, R. H. 2004, \icarus, 170, 131
\bibitem[Emery et al.(2006)]{Emery2006} Emery, J. P., Cruikshank, D. P., \& Van Cleve, J. 2006, \icarus, 182, 496
\bibitem[Encrenaz(2008)]{Encrenaz2008} Encrenaz, T., 2008, \araa, 46, 57
\bibitem[Fayolle et al.(2011)]{Fayolle2011} Fayolle, E. C., {\" O}berg, K. I., Cuppen, H. M., Visser, R., \& Linnartz, H. 2011, \aap, 529, 74
\bibitem[Fornasier et al.(1999)]{Fornasier1999} Fornasier, S., Lazzarin, M., Barbieri, C., \& Barucci, M. A. 1999, \aaps, 135, 65
\bibitem[Fornasier et al.(2014)]{Fornasier2014} Fornasier, S., Lantz, C., Barucci, M. A., \& Lazzarin, M. 2014, \icarus, 233, 163
\bibitem[Fornasier et al.(2016)]{Fornasier2016} Fornasier, S., Mottola, S., Keller, H. U., et al. 2016, Science, 354, 6319
\bibitem[Fraeman et al.(2014)]{Fraeman2014} Fraeman, A. A., Murchie, S. L., Arvidson, R. E., et al. 2014, \icarus, 229, 196
\bibitem[Fujiya et al.(2019)]{Fujiya2019} Fujiya, W., Hoppe, P., Ushikubo, T., et al. 2019, Nat. Ast., 378
\bibitem[Gaffey et al.(1993)]{Gaffey1993} Gaffey, M. J., Burbine, T. H., \& Binzel, R. 1993, Meteoritics, 28, 161
\bibitem[Gartrelle et al.(2021)]{Gartrelle2021} Gartrelle, G. M., Hardersen, P. S., Izawa, M. R. M., \& Nowinski, M. C. 2021, \icarus, 363, 114295
\bibitem[Geake \& Dollfus(1986)]{Geake1986} Geake, J., \& Dollfus, A. 1986, \mnras, 218, 75
\bibitem[Geake \& Geake(1990)]{Geake1990} Geake, J., \& Geake M. 1990, \mnras, 245, 46
\bibitem[Gounelle et al.(2006)]{Gounelle2006} Gounelle, M., Spurn{\'y}, P., \& Bland, P. A. 2006, M\&PS, 41, 135
\bibitem[Gounelle et al.(2008)]{Gounelle2008} Gounelle, M., Morbidelli, A., Band, P. A., et al. 2008, in The solar system beyond Neptune, ed. M. A. Barucci, H. Boehnhardt, D. P. Cruikshank, \& A. Morbidelli (Tucson: The University of Arizona Press), 525
\bibitem[Gourier et al.(2008)]{Gourier2008} Gourier, D., Robert, F., Delpoux, O., et al. 2008, \gca, 72, 1914
\bibitem[Gundlach \& Blum(2013)]{Gundlach2013} Gundlach, B., \& Blum, J. 2013, \mnras, 223, 479
\bibitem[Harris \& Drube(2014)]{Harris2014} Harris, A. W., \& Drube, L. 2014, \apjl, 785, 4
\bibitem[Hasegawa et al.(2022)]{Hasegawa2021} Hasegawa, S., Marsset, M., DeMeo, F. E., et al. 2022, \apjl, 924, L9
\bibitem[Hayano et al. (2010)]{Hayano2010} Hayano, Y., ,Takami, H., Oya, S., et al. 2010, SPIE, 7736, 21
\bibitem[Hiroi et al.(1993)]{Hiroi1993} Hiroi, T., Pieters, C. M., Zolensky, M. E., \& Lipschutz, M. E. 1993, Science, 261, 1016
\bibitem[Hiroi et al.(1996)]{Hiroi1996} Hiroi, T., Zolensky, M. E., Pieters, C. M., \& Lipschutz, M. E. 1996, M\&PS, 31, 321
\bibitem[Hiroi et al.(2001)]{Hiroi2001} Hiroi, T., Zolensky, M. E., \& Pieters, C. M. 2001, Science, 293, 2234
\bibitem[Hiroi et al.(2003)]{Hiroi2003} Hiroi, T., Kanno, A., Nakamura, R., et al. 2003, LPSC, 34, 1425
\bibitem[Hiroi et al.(2005)]{Hiroi2005} Hiroi, T., Tonui, E., Pieters, C. M., et al. 2005, LPSC, 36, 1564
\bibitem[Hiroi et al.(2020)]{Hiroi2020} Hiroi, T., Milliken, R. E., Robertson, K. M., et al. 2020, LPSC, 51, 1043
\bibitem[Howard et al.(2009)]{Howard2009} Howard, K. T., Benedix, G. K., Bland, P. A., \& Cressey, G. 2009, \gca, 73, 4576
\bibitem[Hung et al.(2022)]{Hung2022} Hung, D., Hanu{\v s}, J., Masiero, J. R., \& Tholen, D. J. 2022, PSJ, 3, 56
\bibitem[Ishiguro et al.(2011)]{Ishiguro2011} Ishiguro, M., Hanayama, H., Hasegawa, S., et al. 2011, \apjl, 740, L11
\bibitem[Jedicke et al.(2004)]{Jedicke2004} Jedicke, R., Nesvorn{\'y}, D., Whiteley, R., Ivezi{\'c}, {\v Z}, \& Juri{\'c}, M. 2004, \nat, 429, 275
\bibitem[Johnson \& Fanale(1973)]{Johnson1973} Johnson, T. V., \& Fanale, F. P. 1973, \jgr, 78, 8507
\bibitem[Jones \& Brearley(2006)]{Jones2006} Jones, C. L., \& Brearley, A. J. 2006, \gca, 70, 1040
\bibitem[Kanno et al.(2003)]{Kanno2003} Kanno, A., Hiroi, T., Nakamura, R., et al. 2003, \grl, 30, 1909
\bibitem[Kareta et al.(2021)]{Kareta2021} Kareta, T., Hergenrother, C., Reddy, V., \& Harris, W. M. 2021, PSJ, 2, 31
\bibitem[Khare et al.(1990)]{Khare1990} Khare, B. N., Thompson, W. R., Sagan, C., et al. 1990, First International Conference on Laboratory Research for Planetary Atmospheres (NASA Conference Publication 3077; Washington, DC: NASA), 340
\bibitem[Khare et al.(1993)]{Khare1993} Khare, B. N., Thompson, W. R., Cheng, L., et al. 1993, \icarus, 103, 290
\bibitem[Kimura et al.(2022)]{Kimura2022} Kimura, H., Markkanen, J., Kolokolova, L., 2022, \icarus, 380, 114964
\bibitem[Kiselev et al.(2015)]{Kiselev2015} Kiselev, N., Rosenbush, V., Kolokolova, L., \& Levasseur-Regourd, A. C. 2015, Polarimetry of Stars and Planetary Systems, eds. L. Kolokolova, J. Hough, \& A. Levasseur-Regourd (Cambridge: Cambridge University Press), 379
\bibitem[Kobayashi et al.(2000)]{Kobayashi2000} Kobayashi, N., et al. 2000, IRCS: Infrared Camera and Spectrograph for the Subaru Telescope, Proc. SPIE, vol. 4008, 1056
\bibitem[Kurokawa et al.(2020)]{Kurokawa2020} Kurokawa, H., Shibuya, T., Sekine, Y., et al. 2020, AGU Advances, 3, e2021AV000568
\bibitem[Kwok(2016)]{Kwok2016} Kwok, S. 2016, \araa, 24, 8
\bibitem[Kwon et al.(2021)]{Kwon2021} Kwon, Y. G., Kolokolova, L., Agarwal, J., \& Markkanen, J. 2021, \aap, 650, L7
\bibitem[Lantz et al.(2017)]{Lantz2017} Lantz, C., Brunetto, R., Barucci, M. A., et al. 2017, \icarus, 285, 43
\bibitem[Lantz et al.(2018)]{Lantz2018} Lantz, C., Binzel, R. P., \& DeMeo, F. E. 2018, \icarus, 302, 10
\bibitem[Larsen et al.(2020)]{Larsen2020} Larsen, K. K., Wielandt, D., Schiller, M., Krot, A. N., \& Bizzarro, M. 2020, E\&PSL, 535, 116088
\bibitem[Larson(2010)]{Larson2010} Larson, S. 2010, IAU Circ., 9188, 1
\bibitem[Lazzaro et al.(2004)]{Lazzaro2004} Lazzaro, D., Angeli, C. A., Carvano, J. M., et al. 2004, \icarus, 172, 179
\bibitem[Lebofsky(1980)]{Lebofsky1980} Lebofsky, L. A. 1980, \aj, 85, 573
\bibitem[Lebofsky et al.(1981)]{Lebofsky1981} Lebofsky, L. A., Feierberg, M. A., Tokunaga, A. T., et al. 1981, \icarus, 48, 453
\bibitem[Lebofsky et al.(1986)]{Lebofsky1986} Lebofsky, L. A., Sykes, M. V., Tedesco, E. F., et al. 1986, \icarus, 68, 239
\bibitem[Levison et al.(2009)]{Levison2009} Levison, H. F., Bottke, W. F., Gounelle, M., et al. 2009, \nat, 460, 364
\bibitem[Levison et al.(2020)]{Levison2020} Levison, H., Olkin, C. B., Noll, K. S., et al. 2020, PSJ, 2, 171
\bibitem[Lipschutz et al.(1989)]{Lipschutz1989} Lipschutz, M. E., Gaffey, M. J., \& Pellas, P. 1989, Asteroids II, 740
\bibitem[Lis et al.(2019)]{Lis2019} Lis, D. C., Bockel{\' e}e-Morvan, D., G{\" u}sten, R., et al. 2019, \aap, 625, L5
\bibitem[Lodders(2003)]{Lodders2003} Lodders, K. 2003, \apj, 591, 1220
\bibitem[Lodders(2004)]{Lodders2004} Lodders, K. 2004, \apj, 611, 587
\bibitem[Loeffler \& Prince(2022)]{Loeffler2002} Loeffler, M. J., \& Prince, B. S. 2022, \icarus, 376, 114881
\bibitem[Lord(1992)]{Lord1992} Lord, S. D. 1992, NASA Technical Memorandum 103957
\bibitem[Lumme \& Muinonen(1993)]{Lumme1993} Lumme, K., \& Muinonen, K. 1993, Asteroids, Comets, Meteors, IAU Symp., 160, 194
\bibitem[Lupishko(2022)]{Lupishko2022} Lupishko, D., Ed. 2022. Asteroid Polarimetric Database V2.0. urn:nasa:pds:asteroid\_polarimetric\_database::2.0, NASA Planetary Data System
\bibitem[Mainzer et al.(2019)]{Mainzer2019} Mainzer, A., Bauer, J., Cutri, R., Grav, T., Kramer, E., Masiero, J., Sonnett, S., \& Wright, E., Eds. 2019, NEOWISE Diameters and Albedos V2.0, urn:nasa:pds:neowise\_diameters\_albedos::2.0. NASA Planetary Data System
\bibitem[Marsset et al.(2020)]{Marsset2020} Marsset, M., DeMeo, F. E., Binzel, R. P., et al. 2020, \apjs, 247, 73
\bibitem[Masiero et al.(2011)]{Masiero2011} Masiero, J. R., Mainzer, A. K., Grav, T., et al. 2011, \apj, 741, 68
\bibitem[McAdam et al.(2015)]{McAdam2015} McAdam, M. M., Sunshine, J. M., Howard, K. T., \& McCoy, T. M. 2015, \icarus, 245, 320
\bibitem[McDonough \& Yoshizaki(2021)]{McDonough2021} McDonough, W. F., \& Yoshizaki, T. 2021, Progress in Earth and Planetary Science, 8, 39
\bibitem[McSween et al.(2002)]{McSween2002} McSween Jr., H. Y., Ghosh, A., Grimm, R. E., Wilson, L., \& Young, E. D. 2002, Asteroids III, 559
\bibitem[McSween et al.(2018)]{McSween2018} McSween Jr., H. Y., Emery, J. P., Rivkin, A. S., et al. 2018, M\&PS, 53, 1793
\bibitem[Meier \& Owen(1999)]{Meier1999} Meier, R., \& Owen, T. C. 1999, \ssr, 90, 33
\bibitem[Milliken \& Mustard(2007)]{Milliken2007} Milliken, R. E., \& Mustard, J. R. 2007, \icarus, 189, 574
\bibitem[Miyamoto \& Zolensky(1994)]{Miyamoto1994} Miyamoto, M., \& Zolensky, M. E. 1994, Meteoritics, 29, 849
\bibitem[Poch et al.(2020)]{Poch2020} Poch, O., Istiqomah, I., Quirico, E., et al. 2020, Science, 367, 1212
\bibitem[Prince \& Loeffler(2022)]{Prince2022} Prince, B. S., \& Loeffler, M. J. 2022, \icarus, 372, 114736
\bibitem[Quirico et al.(2016)]{Quirico2016} Quirico, E., Moroz, L. V., Schmitt, B., et al. 2016, \icarus, 272, 32
\bibitem[Remusat et al.(2010)]{Remusat2010} Remusat, L., Guan, Y., Wang, Y., \& Eiler, J. M. 2010, \apj, 713, 1048
\bibitem[Rivkin et al.(2002)]{Rivkin2002} Rivkin, A. S., Howell, E. S., Vilas, F., \& Lebofsky, L. A. 2002, Asteroids III, 235
\bibitem[Rivkin(2012)]{Rivkin2012} Rivkin, A. S. 2012, \icarus, 221, 744
\bibitem[Rivkin et al.(2015a)]{Rivkin2015a} Rivkin, A. S., Thomas, C. A., Howell, E. S., \& Emery, J. P. 2015, \aj, 150, 198
\bibitem[Rivkin et al.(2015b)]{Rivkin2015b} Rivkin, A. S., Campins, H., Emery, J. P., et al. 2015, Asteroids IV, 65
\bibitem[Ross et al.(1969)]{Ross1969} Ross, H. P., Adler, J. E. M., \& Hunt, G. R. 1969, \icarus, 11, 46
\bibitem[Rubin et al.(2007)]{Rubin2007} Rubin, A. E., Trigo-Rodr{\'i}guez, J. M., Huber, H., \& Wasson, J. T. 2007, \gca, 71, 2361
\bibitem[Takir \& Emery(2012)]{Takir2012} Takir, D., \& Emery, J. P. 2012, \icarus, 219, 641
\bibitem[Takir et al.(2013)]{Takir2013} Takir, D., Emery, J. P., McSween, H. Y., et al. 2013, M\&PS, 48, 1618
\bibitem[Takir et al.(2015)]{Takir2015} Takir, D., Emery, J. P., \& McSween, H. Y. 2015, \icarus, 257, 185
\bibitem[Tholen(1984)]{Tholen1984} Tholen, D. J. 1984, Ph.D. thesis, University of Arizona 
\bibitem[Usui et al.(2019)]{Usui2019} Usui, F., Hasegawa, S., Ootsubo, T., \& Onaka, T. 2019, \pasj, 71, 1
\bibitem[van Kooten et al.(2019)]{vanKooten2019} van Kooten, E., Moynier, F., \& Agranier, A. 2019, PNAS, 116, 18860
\bibitem[van Kooten et al.(2021)]{vanKooten2021} van Kooten, E., Schiller, M., Moynier, F., et al. 2021, \apj, 910, 70
\bibitem[Vernazza et al.(2013)]{Vernazza2013} Vernazza, P., Fulvio, D., Brunetto, R., et al. 2013, \icarus, 225, 517
\bibitem[Vernazza et al.(2015)]{Vernazza2015} Vernazza, P., Marsset, M., Beck, P., et al. 2015, \apj, 806, 204
\bibitem[Vernazza et al.(2016)]{Vernazza2016} Vernazza, P., Marsset, M., Beck, P., et al. 2016, \aj, 152, 54
\bibitem[Vilas \& Gaffey(1989)]{Vilas1989} Vilas, F., \& Gaffey, M. J. 1989, Science, 246, 790
\bibitem[Vilas et al.(1994)]{Vilas1994} Vilas, F., Jarvis, K. S., \& Gaffey, M. J. 1994, \icarus, 109, 274
\bibitem[Vilas et al.(2020)]{Vilas2020} Vilas, F., Smith, B.A., McFadden, L.A., et al. 2020, Vilas Asteroid Spectra V1.0. urn:nasa:pds:gbo.ast.vilas.spectra::1.0. NASA Planetary Data System
\bibitem[Vollstaedt et al.(2020)]{Vollstaedt2020} Vollstaedt, H., Mezger, K., \& Alibert, Y. 2020, \apj, 897, 82
\bibitem[Wolff(1975)]{Wolff1975} Wolff, M. 1975, Appl. Opt., 14, 1395
\bibitem[Wood(2005)]{Wood2005} Wood, J. A. 2005, APSC, 341, 953
\bibitem[Wright et al.(2010)]{Wright2010} Wright, E. L., Eisenhardt, P. R. M., Mainzer, A. K., et al. 2010, \aj, 140, 1868
\bibitem[Yang \& Hsieh(2011)]{Yang2011} Yang, B., \& Hsieh, H. 2011, \apjl, 737, L39
\bibitem[Zellner \& Gradie(1976)]{Zellner1976} Zellner, B., \& Gradie, J. 1976, \aj, 81, 262
\bibitem[Zellner et al.(1977)]{Zellner1977} Zellner, B., Leake, M., Lebertre, T., Duseaux, M., \& Dollfus, A. 1977, LPSC, 8, 1091
\bibitem[Zolensky et al.(1989)]{Zolensky1989} Zolensky, M. E., Bourcier, W. L., \& Gooding, J. L. 1989, \icarus, 78, 411
\bibitem[Zolensky et al.(1997)]{Zolensky1997} Zolensky, M. E., Mittlefehldt, D. W., Lipschutz, M. E., et al. 1997, \gca, 61,5099

\end{thebibliography}
\end{document}